\documentclass[useAMS,usenatbib]{mn2e}
\usepackage{mncite}
\usepackage{graphicx}
\usepackage[a4paper]{hyperref}
\usepackage{amssymb}
\usepackage{aas_macros}
\usepackage{subfig}
\usepackage[para]{threeparttable}
\usepackage{color}
\usepackage{longtable}
\usepackage{ulem}

\title[The alignment of WASP-16, WASP-25 and WASP-31]{Rossiter-McLaughlin effect measurements for WASP-16, WASP-25 and WASP-31\thanks{based on observations made using the CORALIE high resolution {\'e}chelle spectrograph mounted on the 1.2\,m \textit{Euler} Swiss Telescope and the HARPS high resolution {\'e}chelle spectrograph mounted on the ESO 3.6\,m (under proposals 084.C-0185 \& 085.C-0393), both at the ESO La Silla observatory.}}
\author[D. J. A. Brown et al]
{
D. J. A. Brown$^{1}$\thanks{E-mail:djab@st-andrews.ac.uk},
A. Collier Cameron$^{1}$,
D. R. Anderson$^{2}$,
B. Enoch$^{1}$,
\newauthor
C. Hellier$^{2}$,
P. F. L. Maxted$^{2}$,
G. R. M. Miller$^{1}$,
D. Pollacco$^{3}$,
D. Queloz$^{4}$,
\newauthor
E. Simpson$^{3}$,
B. Smalley$^{2}$,
A. H. M. J. Triaud$^{4}$,
I. Boisse$^{5}$,
F. Bouchy$^{5,6}$,
\newauthor
M. Gillon$^{7}$,
G. H{\'e}brard$^{5,6}$
\\
$^{1}$ SUPA, School of Physics and Astronomy, University of St Andrews, North Haugh, St Andrews, Fife KY16 9SS, UK.\\
$^{2}$ Astrophysics Group, School of Chemistry and Physics, Keele University, Staffordshire, ST5 5BG, UK.\\
$^{3}$ Astrophysics Research Centre, School of Mathematics \&\ Physics, Queen's University, University Road, Belfast, BT7 1NN, UK.\\
$^{4}$ Observatoire Astronomique de l'Universit{\'e} de Gen{\`e}ve, Chemin des Maillettes 51, CH-1290 Sauverny, Switzerland\\
$^{5}$ Institut d'Astrophysique de Paris, UMR7095 CNRS, Universit{\'e} Pierre \& Marie Curie, 98bis Bd Arago 75014 Paris, France\\
$^{6}$ Observatoire de Haute Provence, CNRS/OAMP, 04870 St Michel l'Observatoire, France\\
$^{7}$ Institut d'Astrophysique et de G{\'e}ophysique, Universit{\'e} de Li{\`e}ge, All{\'e}e du 6 Ao{\^u}t 17, Bat. B5C, Li{\'e}ge 1, Belgium\\
}

\begin{document}

\date{Accepted 2012 March 22}

\pagerange{\pageref{firstpage}--\pageref{lastpage}} \pubyear{2011}

\maketitle

\label{firstpage}

\begin{abstract}
We present new measurements of the Rossiter-McLaughlin (RM) effect for three WASP planetary systems, WASP-16, WASP-25 and WASP-31, from a combined analysis of their complete sets of photometric and spectroscopic data. We find a low amplitude RM effect for WASP-16 ($T{\rm eff}=5700\pm150$\,K), suggesting that the star is a slow rotator and thus of an advanced age, and obtain a projected alignment angle of $\lambda=-4.2^{\circ\,+11.0}_{\,\,-13.9}$. For WASP-25 ($T{\rm eff}=5750\pm100$\,K) we detect a projected spin-orbit angle of $\lambda=14.6^{\circ}\pm6.7$. WASP-31 ($T{\rm eff}=6300\pm100$\,K) is found to be well-aligned, with a projected spin-orbit angle of $\lambda=2.8^{\circ}\pm3.1$. A circular orbit is consistent with the data for all three systems, in agreement with their respective discovery papers. We consider the results for these systems in the context of the ensemble of RM measurements made to date. We find that whilst WASP-16 fits the hypothesis of \citet{winn2010} that `cool' stars ($T{\rm eff}<6250$\,K) are preferentially aligned, WASP-31 has little impact on the proposed trend. We bring the total distribution of the true spin-orbit alignment angle, $\psi$, up to date, noting that recent results have improved the agreement with the theory of \citet{fabrycky2007} at mid-range angles.  We also suggest a new test for judging misalignment using the Bayesian Information Criterion, according to which WASP-25\,b's orbit should be considered to be aligned.
\end{abstract}

\begin{keywords}
planetary systems
--
stars:individual:WASP-16
--
stars:individual:WASP-25
--
stars:individual:WASP-31
--
techniques:radial velocities
\end{keywords}

\section{Introduction}
\label{sec:intro}
As the number of transiting ``hot Jupiters'' known to astronomy has grown, there has been a gradually increasing push towards fully categorising their physical and orbital properties. It is widely presumed that close-in gas giants do not form at the locations in which we observe them, and there are competing theories to describe the process that leads them to their observable orbits.

Migration induced by a protoplanetary disc provides one means by which such a situation can be explained \citep{lin1996}. Since such discs are generally aligned with the host star owing to angular momentum conservation, we would expect that disc migration would preferentially produce well-aligned hot Jupiter systems. Some misaligned planets would not be out of place under this mechanism, being the result of close planet-planet encounters following migration, but we would expect the majority of planets to exhibit spin-orbit alignment.

The Kozai-Lidov mechanism \citep{kozai1962, lidov1962} is the basis of a competing theory for which evidence is mounting. The presence of a third, outer body in a planetary system can excite periodic oscillations in both the eccentricity and inclination of a planetary orbit; inward migration then follows, with tidal friction kicking in as the planet approaches its host, causing the orbit to shrink and circularise \citep{fabrycky2007}. The oscillating inclination that results from Kozai-Lidov interactions produces a continuum of inclinations once the orbits are stable, and thus we would expect the majority of hot Jupiters to exhibit misaligned orbits if the Kozai-Lidov mechanism operates.

It is possible, to some extent, to distinguish between these competing theories through measurement of the spin-orbit alignment angles of hot Jupiter systems. Given the different angular distributions predicted by these theories, building up a significant number of spin-angle measurements is a useful means of determining which mechanism is acting. Unfortunately the true misalignment angle cannot be measured unless a spectroscopic measurement of $v\sin I$ is made, and the stellar rotation period is known. This yields an estimate of the inclination axis to the line-of-sight (e.g. \citet{schlaufman2010}).  The situation is made more difficult by systematic uncertainties in $v\sin I$ measurements, and the sine function, which flattens as it approaches $90^{\circ}$ and therefore only yields useful measurements at low to intermediate inclinations. We are thus currently limited to measuring he projected spin-orbit misalignment angle in the plane of the sky. This is generally measured through the Rossiter-McLaughlin (RM) effect \citep{rossiter1924, mclaughlin1924} which is observable during transit. As the planet transits the approaching limb of the star its spectrum is red-shifted, and when it transits the receding limb its spectrum is blue-shifted. The precise form of the RM anomaly in the radial velocity (RV) curve gives the projected misalignment angle, $\lambda$.

The first observation of the RM effect for a transiting planet was made by \citet{queloz2000}, and since then the number of measurements has increased significantly to a level such that it is possible to begin carrying out analysis of the ensemble of measurements. \citet{fabrycky2009} investigated 11 systems with known values of $\lambda$, deriving two theoretical distributions for $\psi$, the true misalignment angle, using different assumptions about the form of the distribution. They suggested, based on an apparent dual population within their data set, that there might be two routes for planet migration, one producing mostly aligned planets and the other producing misaligned planets.

One early indication of a pattern was that misaligned planets tended to be high mass and on eccentric orbits \citep{johnson2009}. Subsequent observations have often countered this initial trend (for example HAT-P-7 \citep{winn2009b, narita2009}), but high mass ($M_P>4$\,$M{\rm Jup}$) planets do appear to have a different obliquity distribution \citep{hebrard2011}. Of the 6 planets in this category with measured misalignment angles four are misaligned, but none have $|\lambda|>50^{\circ}$. More observations of high mass planets are needed before we can be certain that this is not merely an artefact of small-number statistics however.

One of the more intriguing suggestions was put forward by \citet{winn2010} (hereafter \defcitealias{winn2010}{W10}\citetalias{winn2010}), who speculated that the division into aligned and misaligned planets might be dependent on the effective temperature of the host star. Using a larger sample of 19 systems with known $\lambda$, they found that the misaligned systems were preferentially hotter than the aligned examples, with a critical temperature of $T{\rm eff}\approx6250$\,K dividing the two populations. One explanation put forward for this was the tidal realignment of planets around `cool' stars, with the equivalent process around `hot' stars being suppressed owing to their lack of a convective envelope. \citetalias{winn2010} further conjecture that the current $\psi$ distribution could be completely explained by a migration mechanism driven by a combination of Kozai-Lidov oscillations and planet-planet scattering, without the need to invoke disc migration.

\citet{triaud2010} (hereafter \defcitealias{triaud2010}{T10}\citetalias{triaud2010}) added 6 planets to the ensemble of known RM measurements. Calculating individual $\psi$ distributions for each planet based on the assumption that stellar rotation axes are randomly oriented on the sky, they produced a total distribution for the ensemble of planets, finding that it matched the theoretical distribution of \citet{fabrycky2007} for Kozai-Lidov mechanism dominated migration, further implying that disc migration might be superfluous to requirements for explaining the presence of hot Jupiters.

\begin{table*}
	\caption{System parameters for the three WASP planetary systems for which we evaluate the Rossiter-McLaughlin effect. Parameters for WASP-16 were taken from \citet{lister2009}. Parameters for WASP-25 were taken from \citet{enoch2011}. Parameters for WASP-31 were taken from \citet{anderson2011}. $v\sin I$ and macroturbulence values have been updated through spectroscopic analysis of the new HARPS data using the \citet{bruntt2010} calibration.}
	\label{tab:planets}
	\begin{tabular}{lllll}
		\hline \\
		Parameter    & Unit          & WASP-16                      & WASP-25                & WASP-31 \\
		\hline \\
		$M_*$        & $M_\odot$     & $1.022^{+0.074}_{-0.129}$    & $1.00\pm0.03$          & $1.161\pm0.026$ \\[2pt]
		$R_*$        & $R_\odot$     & $0.946^{+0.057}_{-0.052}$    & $0.92\pm0.04$          & $1.241\pm0.039$ \\[2pt]
		$T{\rm eff}$ & K             & $5700\pm150$                 & $5750\pm100$           & $6300\pm100$ \\[2pt]
		$v\sin I$    & km\,s$^{-1}$  & $2.3\pm0.4$                  & $2.6\pm0.4$            & $8.1\pm0.5$ \\[2pt]
		macroturbulence & km\,s$^{-1}$  & $2.3$ & $2.4$ & $4.2$ \\[2pt]
		$M_p$        & $M{\rm Jup}$  & $0.855^{+0.043}_{-0.076}$    & $0.58\pm0.04$          & $0.478\pm0.030$ \\[2pt]
		$R_p$        & $R{\rm Jup}$  & $1.008^{+0.083}_{-0.060}$    & $1.22^{+0.06}_{-0.05}$ & $1.537\pm0.060$ \\[2pt]
		$P$          & days          & $3.11860\pm0.00001$          & $3.764825\pm0.000005$  & $3.405909\pm0.000005$ \\[2pt]
		$a$          & AU            & $0.0421^{+0.0010}_{-0.0019}$ & $0.0473\pm0.0004$      & $0.04657\pm0.00034$ \\[2pt]
		$e$          &               & $0$(adopted)                 & $0$(adopted)           & $0$(adopted) \\[2pt]
		$i$          & $^{\circ}$    & $85.22^{+0.27}_{-0.43}$      & $88.0\pm0.5$           & $84.54\pm0.27$ \\[2pt]
		\hline \\
	\end{tabular}
\end{table*}

Here we present measurements of the RM angle for three more planets from the Wide Angle Search for transiting Planets (WASP) \citep{pollacco2006}, WASP-16\,b, WASP-25\,b and WASP-31\,b, and investigate how they modify the ensemble results and conclusions discussed above. In section\,\ref{sec:obs} we give details of our observations, and in section\,\ref{sec:data} we discuss the analytical methods used to determine the misalignment angles. In section\,\ref{sec:results} we report on the results of our analysis for the individual systems. In section\,\ref{sec:discuss} we discuss the implications of our results for previously observed trends. Finally, in section\,\ref{sec:misalign}, we take another look at the question of alignment, presenting a new test for planetary orbit misalignment.

\section{Observations}
\label{sec:obs}
Radial velocity data for all three planetary systems were obtained using the CORALIE high precision {\'e}chelle spectrograph \citep{queloz2000proc} mounted on the Swiss 1.2\,m Euler telescope, and with the HARPS high precision {\'e}chelle spectrograph \citep{mayor2003} mounted on the 3.6\,m ESO telescope at La Silla. Data from CORALIE were used primarily to constrain the presence of a long-term trend in radial velocity that might be indicative of a third body in the system, whilst HARPS was used to monitor the radial velocity before, during and after a specific transit event. Two data points were obtained the night before the transit, and for at least one night following the transit; on the night of the transit observations were started 90 minutes prior to the predicted start of transit and continued until 90 minutes after its predicted conclusion.

\subsection{WASP-16}
\label{sec:W16obs}
WASP-16 was observed using CORALIE between 2008 March 10 and 2009 June 3, on an ad-hoc basis. One datum was also acquired on 2010 July 14 to retest the hypothesis of a long-term radial velocity trend. The transit observed with HARPS occurred on the night of 2010 March 21; 32 data points were acquired over the duration of the night. This transit observation was affected by cloud cover, so an additional transit was observed on the night of 2011 May 12, producing a further 28 RV measurements. Further measurements were made on the days surrounding this transit as well (see journal of observations, Tables\,\ref{tab:W16coralie}, \ref{tab:W16harps} and \ref{tab:W16harps2}).

Details of the photometric observations of WASP-16 are given in \citet{lister2009}.

\subsection{WASP-25}
\label{sec:W25obs}
HARPS observed the transit taking place on the night of 2008 April 11. 44 observations were made that night, with additional data acquired on adjacent nights (see the journal of observations, Tables\,\ref{tab:W25coralie} and \ref{tab:W25harps}). The system was observed using CORALIE between 2008 December 29 and 2009 June 28, with observations made at irregular intervals between these dates.

\citet{enoch2011} describe the photometric observations that were made of WASP-25.

\subsection{WASP-31}
\label{sec:W31obs}
WASP-31 was observed using CORALIE between 2009 January 4 and 2010 May 18 during several short runs. HARPS was used to observe a full transit on the night of 2010 April 15, with 17 data points obtained. Additional observations were made on adjacent nights (see the journal of observations, Tables\,\ref{tab:W31coralie} and \ref{tab:W31harps}).

The photometric observations for WASP-31 are discussed in \citet{anderson2011}.

\section{Data Analysis}
\label{sec:data}
Our analysis mirrors that of \citetalias{triaud2010}, using the complete set of photometric and spectroscopic data for the objects that we investigate in order to fully account for parameter correlations. We use an adapted version of the code described in \citet{cameron2007}, fitting models of the photometric transit, the Keplerian RV and the RM effect to the system data. The fit of our model is refined using a Markov Chain Monte Carlo (MCMC) technique to minimize the $\chi^2$ statistic, and to explore the parameter space using the jump parameters $T_0$ (epoch of mid-transit), $P$ (orbital period), $W$ (transit width), $b$ (impact parameter), $\gamma$ velocity, $\dot{\gamma}$, $K$ (RV semi-amplitude), $T{\rm eff}$ (stellar effective temperature), $[Fe/H]$ (metallicity), $\sqrt{e}\cos\omega$, $\sqrt{e}\sin\omega$, $\sqrt{v\sin I}\cos\lambda$ and $\sqrt{v\sin I}\sin\lambda$. We use a burn-in phase of 2000 steps, with burn-in judged to be complete when $\chi^2$ becomes greater than the median of all previous values \citep{knutson2008}. A minimum burn-in length of 500 steps is applied to ensure that burn-in is truly complete. Once this initial phase is over we use a further 100 steps to recalculate the parameter jump lengths before beginning the real Markov Chain of 10000 accepted steps; with the acceptance rate of 25\,percent recommended for the Metropolis-Hastings algorithm \citep{tegmark2004} this gives an effective chain length of 40000 steps. Our set of final parameters is taken to be the median of the Markov chain, with the $1\sigma$ error bars calculated from the values that encompass the central 68.3\,percent of the accepted steps. We account for limb darkening using a non-linear treatment based on the tables of \citet{claret2000}, interpolating the coefficients at each step in the chain. 

The inclusion of the photometric data is an important point. Although we fit the RM effect to the radial velocity data, the transit width and depth, as well as the impact parameter, can be determined from the photometric transit. These parameters have a role to play in the characterisation of the form of the RM anomaly. The transit width helps to determine the duration of the anomaly whilst the depth gives the planetary and stellar radii. The radii and impact parameter in turn help to determine $v\sin I$, upon which the amplitude of the anomaly depends \citep{queloz2000}. Although characterisation of the RM effect can be carried out using the spectroscopic data alone, by taking the photometric data into account in this way we ensure consistency across the full set of system parameters. To account for stellar jitter we initially assign a value of $1$\,m\,s$^{-1}$, below the level of precision of the spectrographs used for this work, which we added in quadrature to the in-transit photometric data.

We separate our RV data by instrument, and within those distinctions also treat spectroscopic data taken on nights featuring planetary transits as separate datasets. Our model for the orbital RV signature treats the sets of data as independent, producing individual offsets and radial velocity trends for each one. The reported solution is that for the set of RV data covering the greatest phase range. For completeness, we also repeated our analysis using \textit{only} RV data taken during nights that featured a transit event, but found little to distinguish them from our analysis of the the full set of data.

For our RM model we use the analytic formula of \citet{hirano2011b}. This method requires prior knowledge of several broadening coefficients, specifically the macroturbulence, for which our estimates are noted in Table\,\ref{tab:planets}, and the Lorentzian ($\gamma$) and Gaussian ($\beta$) spectral line dispersions. The line dispersions were dictated by our use of the HARPS instrument, which has a spectral resolution of $R=115000$, implying an instrumental Gaussian dispersion of $2.61$\,km\,s$^{-1}$. This was combined with the intrinsic Doppler linewidth, including appropriate thermal and turbulent motion for each star, to obtain values of $\beta=3.1$\,km\,s$^{-1}$ for WASP-16 and WASP-25, and $\beta=3.3$\,km\,s$^{-1}$ for WASP-31. We assumed $\gamma=0.9$\,km\,s$^{-1}$ in line with \citeauthor{hirano2011b}, and also assumed that the coefficient of differential rotation, $\alpha=0$. WASP-16 and WASP-25 are both slow rotators, and whilst WASP-31 should be considered a moderately fast rotator, without knowledge of the inclination of the stellar rotation axis it is difficult to place a value of $\alpha$.

We apply several Bayesian priors to $\chi^2$ to account for previously known information: a prior on the eccentricity, allowing for the forcing of circular solutions; a prior on the spectroscopic $v\sin I$, using updated values of $v\sin I$ derived from the newly acquired HARPS spectra and the macroturbulence calibration of \citet{bruntt2010}, and a prior enforcing a main sequence (MS) mass-radius relationship. This MS prior is based on that discussed in \citet{cameron2007}, but is only applied to the stellar radius. The stellar mass is estimated using the calibration of \citet{enoch2010}.

To distinguish between models that use different combinations of priors we minimize the reduced $\chi^2$ for the spectroscopic data; in cases where there is little to choose between the different sets of input conditions we gravitate towards the model with the fewest free parameters. In what follows we refer to $\chi^2$ as the combined $\chi^2$ for the complete data set, $\chi^2_{RV}$ as the value for the spectroscopic RV data only, and $\chi^2_{red}$ as the reduced $\chi^2$ for the spectroscopic data alone. Note also that we refer to the projected spin-orbit misalignment angle as $\lambda$, as is more common in the literature, not $\beta$ as used by \citetalias{triaud2010} (strictly $\lambda=-\beta$).

\section{Rossiter-McLaughlin results}
\label{sec:results}
\subsection{WASP-16}
\label{sec:W16res}
WASP-16b \citep{lister2009} (hereafter \defcitealias{lister2009}{L09}\citetalias{lister2009}) is a close Jupiter analog orbiting a Solar-type star with a period of $3.12$ days. The planet is somewhat less massive than Jupiter but of comparable radius, whilst the host star is similar in mass, radius and metallicity to the Sun, but exhibits significant lithium depletion. Our updated spectroscopic analysis using the HARPS spectra yields a projected stellar rotation velocity of $v\sin I=2.3\pm0.4$\,km\,s$^{-1}$.

Our original estimate of stellar jitter produced fits with $\chi^2\approx1.6$, leading us to re-estimate the jitter following \citet{wright2005}. We calculated line strengths for the calcium H and K emission lines in each of the HARPS spectra, and used these to estimate values for the chromospheric activity metric S. These were then calibrated against the Mount Wilson sample (see e.g. \citet{baliunas1995}), and absolute magnitudes of the stars were calculated using \citet{gray1992}. We eventually adopted the 20th percentile value of $3.6$\,m\,s$^{-1}$ as a conservative estimate of the jitter.

Removing the requirement for the system to obey a main sequence mass-radius relationship (equation 6 in \citet{cameron2007}) produced changes of between $0$ and $2$\,percent in the stellar mass and radius, leading to increases in the stellar density of between $1$ and $4$\,percent, for no discernible improvement in fit. Comparing impact parameter values, we find that we obtain an average value of $\bar{b}=0.83^{+0.03}_{-0.04}$ for the cases both with and without the MS prior active. The parameter S \citep{cameron2007},
\begin{equation}
S=-2\ln P(M_*,R_* ) = \frac{R_*-R_0}{\sigma_R^2},
\label{eq:S}
\end{equation}
used to measure the discrepancy between the stellar radius from the (J-H) colour and that returned by the MCMC algorithm, increases from an average of $0.17$ to $0.34$ when the prior is removed, a relatively small increase as suggested by the modest changes in stellar parameters. We therefore find little to distinguish between the cases with the MS requirement applied, and those with the stellar radius freely varying, and choose not to apply this prior in our final solution.

Adding a long-term, linear RV trend produced no improvement in $\chi^2_{red}$, and with a magnitude of $|\dot{\gamma}|<3$\,m\,s$^{-1}$\,yr$^{-1}$ we disregard the possibility that there is a such a trend in the spectroscopic data. Adding a prior on the spectroscopic $v\sin I$ similarly gave almost no difference in the quality of the fit obtained. For most combinations of priors our analysis returned $v\sin I\approx1.2\pm0.3$\,km\,s$^{-1}$, significantly slower than the spectroscopic value.

Allowing the eccentricity to float again led to no significant improvement in the fit, and all of the values of $e$ returned by our various combinations of priors were consistent with $e=0$ to within $2\sigma$. We tested these small eccentricity values using equation\,27 of \citet{lucy1971}, which adopts a null hypothesis of a circular orbit and considers an orbit to be eccentric if this is rejected at the 5\,percent significance level. This F-test indicated that none of the eccentricities were significant, and thus that a circular orbit is favoured.

\begin{figure*}
	\subfloat{
		\includegraphics[width=0.48\textwidth]{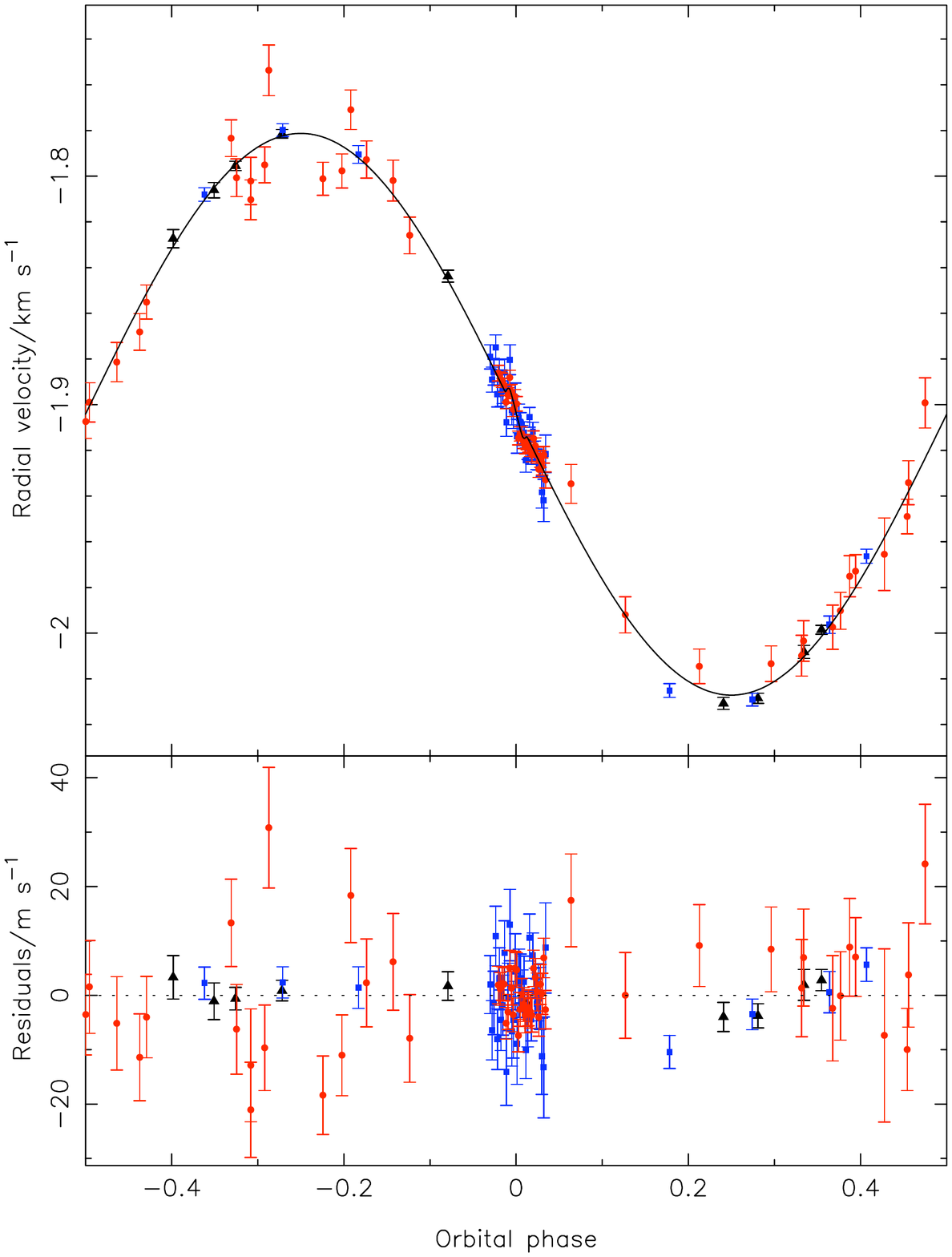}
		\label{fig:W16RV}}
	\subfloat{
		\includegraphics[width=0.48\textwidth]{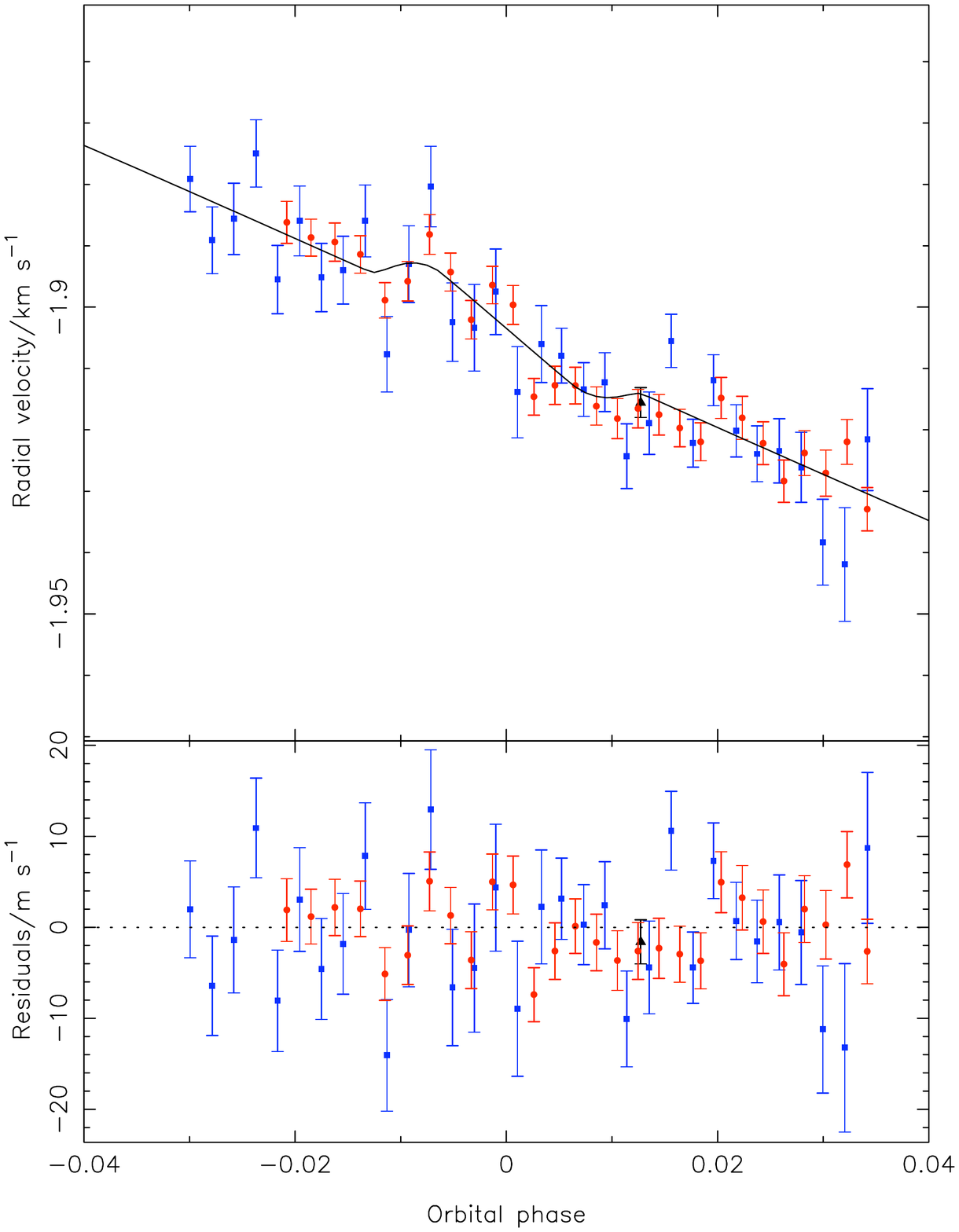}
		\label{fig:W16RM}}
	\caption{Results from the fit to the data for WASP-16 using $e=0$, no long-term radial velocity trend, no prior on the spectroscopic $v\sin I$, and without forcing the mass-radius relationship. Black, filled triangles represent data from CORALIE. Blue, filled squares represent data from the first HARPS run. Red, filled circles represent data from the second HARPS run. The best-fitting model is plotted as a solid black line. \textit{Top left:} Complete radial velocity reflex motion curve. \textit{Bottom left:} Residuals from theRV fit, exhibiting no correlation with phase. \textit{Top right:} Close up of the transit region from the radial velocity curve showing the RM effect, along with the residuals. \textit{Bottom right:} Residuals for the RV data within the RM window.}
	\label{fig:W16res}
\end{figure*}

\begin{figure*}
	\subfloat{
		\includegraphics[width=0.48\textwidth]{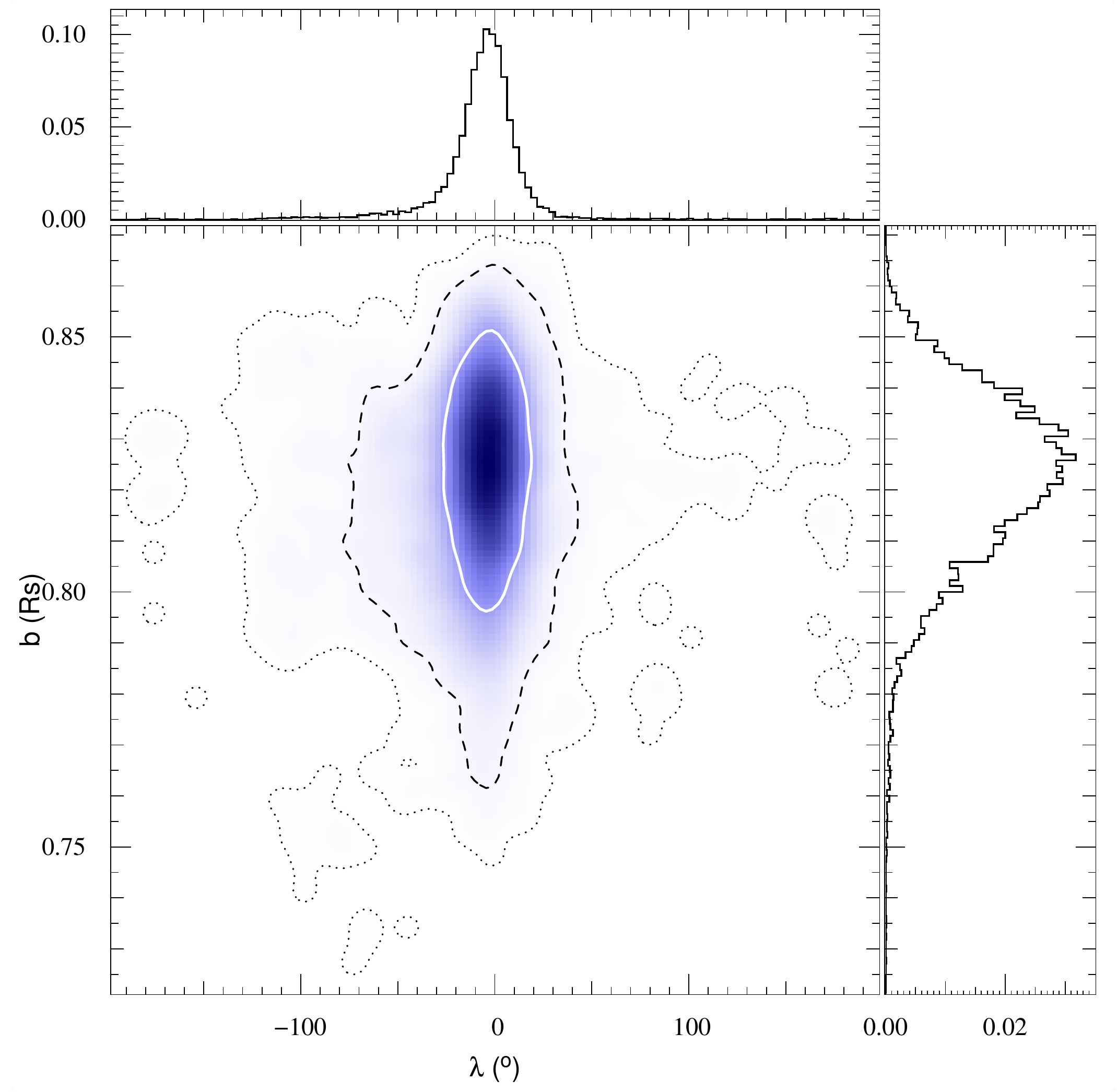}
		\label{fig:W16prob2}}
	\subfloat{
		\includegraphics[width=0.48\textwidth]{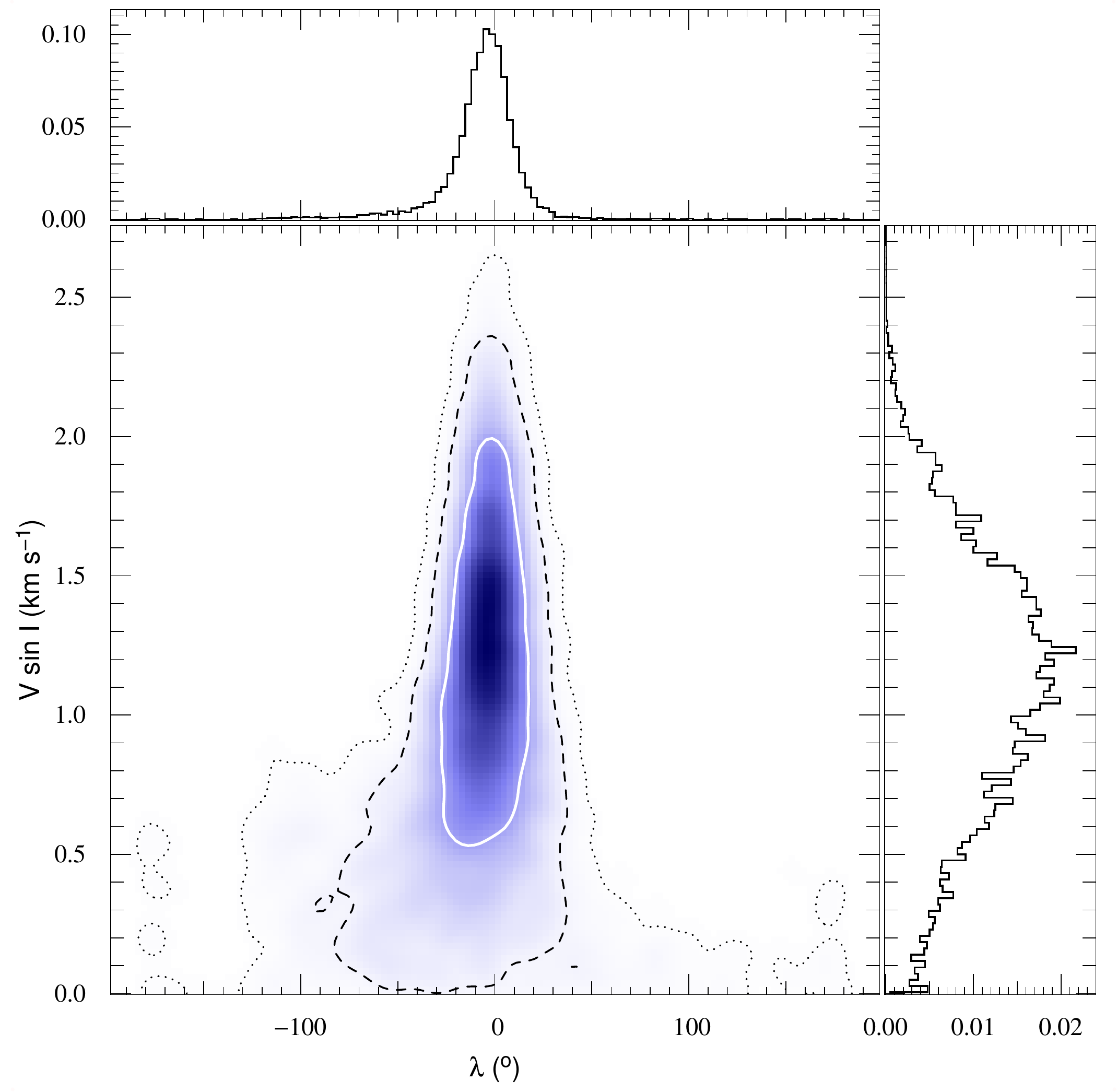}
		\label{fig:W16prob5}}
	\caption{Posterior probability distributions derived from the Markov chains, for the fit to WASP-16 described in Fig.\,\ref{fig:W16res}. The white contours mark the $62.87$\,percent confidence regions, the black, dashed contours the $95.45$\,percent confidence regions, and the black, dotted contours the $99.73$\,percent confidence regions. Marginalised, 1D distributions are displayed in the side panels. \textit{Left:} $b$ and $\lambda$. \textit{Right:} $v\sin I$ and $\lambda$. This distribution has a triangular shape, and $\lambda=0$ falls within the central body of the distribution. Both distributions have poorly constrained $99.73$\,percent confidence regions, and show a slight bias towards negative values of $\lambda$.}
	\label{fig:W16prob}
\end{figure*}

We therefore adopt as our conclusive solution the case without the MS prior active, with no prior on $v\sin I$, no long-term trend in velocity and a circular orbit, but we stress that changing the priors had little impact on the parameter values returned by the MCMC algorithm. Our adopted solution returns values of $\lambda=-4.2^{\circ\,+11.0}_{\,\,13.9}$ and $v\sin I=1.2^{+0.4}_{-0.5}$\,km\,s$^{-1}$; this is significantly slower than the spectroscopic value of $v\sin I$ that we obtained from spectral analysis. However an alternative analysis of the HARPS spectra using the calibration of \citet{gray2008} provides an estimate of $v\sin I=1.2\pm0.5$, in good agreement with the value that we found from our model. Our solution also indicates a high impact parameter of $0.82^{+0.01}_{-0.02}$ that reduces the likelihood of a degeneracy developing between $\lambda$ and $v\sin I$. Examination of Fig.\,\ref{fig:W16prob5} highlights this, with a triangular distribution that is centred close to $\lambda=0^{\circ}$. The main section of this distribution lies within the limits $|\lambda|<20^{\circ}$, providing further evidence for the well-aligned system that was suggested by our best-fitting RM angle. From \citetalias{lister2009} we note that the host star has $T{\rm eff}=5700\pm150$\,K, which places it in the `cool' category of \citetalias{winn2010}; an aligned orbit therefore fits their hypothesis quite nicely.

\begin{table*}
	\caption{A comparison of the $\chi^2$ and $\chi^2_{red}$ values for WASP-16 for each combination of Bayesian priors. All values of $\chi^2$ include the Bayesian penalties applicable for that combination of priors.}
	\label{tab:W16res}
	\begin{tabular}{lllllllll}
	\hline \\
	$v\sin I$ prior & MS prior & $\dot{\gamma}$/ms$^{-1}$yr$^{-1}$ & eccentricity                             & $v\sin I$/km\,s$^{-1}$ & $\lambda$/$^{\circ}$     & $\chi^2$                & $\chi^2_{RV}$ & $\chi^2_{red}$ \\
	\hline \\
	off                     & off            & $0$                                                          & $0.009^{+0.010}_{-0.006}$ & $1.2\pm0.4$                 & $-2.1^{+10.5}_{-11.0}$ & $12915\pm161$ & $100\pm14$    & $0.9\pm0.1$ \\[2pt]
	off                     & on            & $0$                                                          & $0.004^{+0.006}_{-0.002}$ & $1.2^{+0.4}_{-0.5}$    & $-2.8^{+10.7}_{-11.1}$  & $12917\pm161$ & $102\pm14$    & $0.9\pm0.1$ \\[2pt]
	off                     & off            & $1.0^{+0.8}_{-0.8}$                             & $0.011^{+0.009}_{-0.007}$ & $1.0^{+0.5}_{-0.6}$     & $-2.5^{+13.6}_{-16.6}$  & $12912\pm161$ & $99\pm14$    & $0.9\pm0.1$ \\[2pt]
	off                     & on            & $0.6^{+0.5}_{-0.3}$                             & $0.007^{+0.007}_{-0.005}$ & $1.1^{+0.4}_{-0.6}$     & $-3.6^{+10.9}_{-14.8}$  & $12911\pm161$ & $99\pm14$      & $0.9\pm0.1$ \\[2pt]
	off                     & off            & $0$                                                          & $0$                                           & $1.1^{+0.5}_{-0.6}$    & $-6.7^{+11.7}_{-19.2}$  & $12917\pm161$ & $103\pm14$   & $1.0\pm0.1$ \\[2pt]
	off                     & on            & $0$                                                          & $0$                                           & $1.2^{+0.4}_{-0.5}$    & $-4.2^{+11.0}_{-13.9}$  & $12916\pm161$ & $103\pm14$    & $1.0\pm0.1$ \\[2pt]
	off                     & off            & $0.1\pm0.1$                                          & $0$                                           & $1.1^{+0.5}_{-0.6}$     & $-5.8^{+10.6}_{-14.5}$  & $12917\pm161$ & $102\pm14$    & $0.9\pm0.1$ \\[2pt]
	off                     & on            & $0.9^{+1.0}_{-0.9}$                              & $0$                                           & $1.2\pm0.5$                 & $-6.0^{+10.3}_{-15.4}$  & $12911\pm161$ & $102\pm14$    & $0.9\pm0.1$ \\[2pt]
	$2.3\pm0.4$  & off             & $0$                                                          & $0.011^{+0.009}_{-0.008}$ & $1.2\pm0.3$                 & $-1.8^{+11.0}_{-11.2}$  & $12910\pm161$ & $100\pm14$    & $0.9\pm0.1$ \\[2pt]
	$2.3\pm0.4$  & on             & $0$                                                          & $0.012^{+0.009}_{-0.007}$ & $1.2\pm0.2$                 & $-2.3^{+10.5}_{-11.7}$  & $12914\pm161$ & $98\pm14$    & $0.9\pm0.1$ \\[2pt]
	$2.3\pm0.4$  & off             & $0.1\pm0.1$                                          & $0.010^{+0.009}_{-0.007}$ & $1.2\pm0.3$                  & $-3.6^{+11.7}_{-11.3}$  & $12916\pm161$ & $101\pm14$    & $0.9\pm0.1$ \\[2pt]
	$2.3\pm0.4$ & on              & $0.7^{+0.7}_{-0.8}$                              & $0.011^{+0.009}_{-0.007}$ & $1.2\pm0.2$                 & $-2.9^{+9.9}_{-9.0}$       & $12912\pm161$ & $99\pm14$    & $0.9\pm0.1$ \\[2pt]
	$2.3\pm0.4$ & off              & $0$                                                          & $0$                                           & $1.2\pm0.3$                 & $-4.9^{+10.0}_{-11.0}$   & $12912\pm161$ & $102\pm14$    & $0.9\pm0.1$ \\[2pt]
	$2.3\pm0.4$ & on             & $0$                                                           & $0$                                           & $1.2\pm0.3$                 & $-4.8^{+9.6}_{-10.2}$     & $12919\pm161$ & $104\pm14$    & $1.0\pm0.1$ \\[2pt]
	$2.3\pm0.4$ & off             & $2.1^{+3.3}_{-1.8}$                               & $0$                                           & $1.1^{+0.3}_{-0.4}$     & $-5.6^{+10.0}_{-12.9}$   & $12916\pm161$ & $101\pm14$    & $0.9\pm0.1$ \\[2pt]
	$2.3\pm0.4$ & on             & $-0.6^{+1.6}_{-1.3}$                             & $0$                                           & $1.1\pm0.4$                  & $-5.7^{+11.4}_{-12.5}$   & $12917\pm161$ & $103\pm14$    & $0.9\pm0.1$ \\[2pt]
	\hline \\
	\end{tabular}
\end{table*}

As previously noted, the amplitude of the RM anomaly for WASP-16 is quite small. The aligned nature of the system suggests that this can be put down to the star being an old, slowly rotating star, which  would be consistent with the age estimate reported by \citetalias{lister2009}, which suggests an age $>5$\,Gyr based on a lack of detectable lithium. A second possible explanation could be that we are in fact viewing the host star almost pole-on, which could still be consistent with an orbit that is aligned in the plane of the sky. This would lead to a low projected rotation velocity, and a transit across the pole of the star would have a small RM amplitude, as observed here. The minimum stellar inclination is limited by the observed lithium depletion\footnote{The abundance of lithium gives us a minimum age, as stated. If we assume that gyrochronology is applicable, then this provides a maximum true stellar rotation velocity. This in turn allows us to use the detected $v\sin I$ to calculate the minimum possible stellar inclination.}, but such a structure would imply a younger age for the star owing to the rapid true stellar rotation. Interestingly isochronal analysis in \citetalias{lister2009} implies an age of $2.3^{+5.8}_{-2.2}$ Gyr, lower than the limit implied by the lithium depletion. However new isochronal fits, using our results and a range of stellar models, returned ages of $4.7^{+3.3}_{-4.3}$\,Gyr (Padova models; \citet{marigo2008}), $4.8^{+1.2}_{-3.3}$\,Gyr (Yonsei-Yale models; \citet{demarque2004}), $6.0^{+5.0}_{-4.0}$\,Gyr (Teramo models; \citet{pietrinferni2004}) and $5.0^{+4.9}_{-3.8}$\,Gyr (VRSS models; \citet{vandenberg2006}). These ages further support the case for a slowly rotating host star, and are consistent with the star's observed lithium abundance.

Careful analysis of the HARPS spectra allowed us to measure the chromospheric Ca II H \& K emission. We find that $\log(R'_{HK})=-5.10\pm0.15$, indicating a low level of chromospheric activity. This rules out the possibility that the star is misaligned along the line-of-sight, as we would expect much greater calcium emission from a young, rapidly rotating star. We note that this agrees with the work of \citet{schlaufman2010}, who finds no evidence for line-of-sight misalignment in the WASP-16 system. Following \citet{watson2010} we calculate $P_{rot}=30.2^{+4.7}_{-3.8}$\,days, which implies an age of $3.8^{+1.2}_{-0.8}$\,Gyr for WASP-16 according to the gyrochronology method of \citet{barnes2007} using the updated coefficients of \citet{meibom2009} and \citet{james2010}. 
A recent reanalysis of the WASP-1 and WASP-2 systems \citep{albrecht2011} highlighted the fact that in systems with low amplitude, low S/N RM anomalies, the angles reported tend towards $0^{\circ}$ and $180^{\circ}$ owing to the greater probability density in the distribution for $\lambda$. The same study cautions readers against drawing strong conclusions of alignment in such cases. Our data for WASP-16 certainly show some of the characteristics discussed in the \citeauthor{albrecht2011} study, and we have indeed found a well-aligned system with $\lambda$ close to 0. 

However there are other methods by which the alignment angle of a planetary orbit can be deduced. Doppler tomography is an established method for mapping velocity variations in binary stars (e.g. \citet{albrecht2007,albrecht2009}), but its application to transiting exoplanets is in its infancy. The technique has, to date, been used to study HD189733 \citep{cameron2010}, WASP-33 \citep{cameron2010b} and WASP-3 \citep{miller2010}, and is best suited to analysing hot, rapidly rotating exoplanet host stars. WASP-16\,A exhibits neither of these attributes, but analysis is ongoing (Miller et al., \textit{in prep}) and indications are that it gives similar results for the obliquity angle of this system. An independent detection of the RM effect, also suggesting alignment, was announced at IAU Symposium 276 by Winn, and we look forward to the published results with interest.

\subsection{WASP-25}
\label{sec:W25res}
WASP-25b \citep{enoch2011} (hereafter \defcitealias{enoch2011}{E11}\citetalias{enoch2011}) is a significantly bloated, sub-Jupiter mass planet orbiting a solar-type, somewhat metal-poor host star with an orbital period of $3.76$ days. A full set of results from our analysis is displayed in Table\,\ref{tab:W25res}. One RV measurement was found to lie at $3\,\sigma$ from the best-fitting model, and to be consistent with the out of transit RV curve. This datum was omitted from our analysis, and will be discussed further later.

\begin{table*}
	\caption{A comparison of the $\chi^2$ and $\chi^2_{red}$ values for WASP-25 for each combination of Bayesian priors. All values of $\chi^2$ include the Bayesian penalties applicable for that combination of priors.}
	\label{tab:W25res}
	\begin{tabular}{lllllllll}
	\hline \\
	$v\sin I$ prior & MS prior & $\dot{\gamma}$/ms$^{-1}$yr$^{-1}$ & eccentricity                              & $v\sin I$/km\,s$^{-1}$ & $\lambda$/$^{\circ}$       & $\chi^2$                & $\chi^2_{RV}$ & $\chi^2_{red}$ \\
	\hline \\
	off                     & off            & $0$                                                           & $0.011^{+0.010}_{-0.008}$ & $2.8\pm0.3$                 & $17.9^{+9.8}_{-8.6}$    & $14200\pm169$ & $104\pm14$     & $1.3\pm0.2$ \\[2pt]
	off                     & on           & $0$                                                            & $0.013^{+0.013}_{-0.009}$ & $2.8\pm0.3$                 & $15.9^{+7.5}_{-7.3}$      & $14195\pm168$ & $103\pm14$     & $1.3\pm0.2$ \\[2pt]
	off                     & off           & $103.8^{+25.5}_{-29.6}$                       & $0.013^{+0.014}_{-0.009}$ & $2.9\pm0.3$                 & $16.8^{+9.5}_{-9.4}$      & $14184\pm168$ & $90\pm13$       & $1.1\pm0.2$ \\[2pt]
	off                     & on           & $-10.3^{+13.6}_{-10.1}$                        & $0.011^{+0.013}_{-0.008}$ & $2.8\pm0.3$                & $14.9^{+6.6}_{-7.1}$      & $14197\pm169$ & $101\pm14$       & $1.3\pm0.2$ \\[2pt]
	off                     & off           & $0$                                                             & $0$                                           & $2.9\pm0.3$                 & $14.6\pm6.7$                 & $14200\pm169$ & $104\pm14$     & $1.3\pm0.2$ \\[2pt]
	off                     & on           & $0$                                                             & $0$                                           & $2.9\pm0.3$                 & $17.0^{+8.5}_{-8.1}$    & $14199\pm169$ & $103\pm14$     & $1.3\pm0.2$ \\[2pt]
	off                     & off           & $96.1^{+28.7}_{-26.6}$                          & $0$                                           & $2.8\pm0.3$                & $18.8^{+10.1}_{-8.6}$    & $14189\pm168$ & $91\pm13$    & $1.1\pm0.2$ \\[2pt]
	off                     & on           & $2.4^{+0.4}_{-0.3}$                                 & $0$                                           & $2.8\pm0.2$                 & $12.7^{+8.4}_{-5.7}$      & $13754\pm166$ & $103\pm14$    & $1.3\pm0.2$ \\[2pt]
	$2.6\pm0.4$ & off             & $0$                                                            & $0.013^{+0.014}_{-0.009}$  & $2.8\pm0.2$                & $15.6^{+8.9}_{-8.4}$      & $14194\pm168$ & $103\pm14$     & $1.3\pm0.2$ \\[2pt]
	$2.6\pm0.4$ & on             & $0$                                                            & $0.011^{+0.011}_{-0.008}$  & $2.8\pm0.2$                 & $14.5^{+7.6}_{-6.7}$      & $14200\pm169$ & $104\pm14$    & $1.3\pm0.2$ \\[2pt]
	$2.6\pm0.4$ & off             & $100.4^{+28.6}_{-28.4}$                        & $0.013^{+0.014}_{-0.009}$ & $2.8\pm0.2$                 & $16.8^{+9.2}_{-9.0}$     & $14183\pm168$ & $90\pm13$    & $1.1\pm0.2$ \\[2pt]
	$2.6\pm0.4$ & on             & $97.1^{+28.0}_{-25.8}$                          & $0.011^{+0.013}_{-0.008}$  & $2.8\pm0.2$                 & $15.3^{+7.5}_{-6.6}$     & $14187\pm168$ & $91\pm14$      & $1.1\pm0.2$ \\[2pt]
	$2.6\pm0.4$ & off             & $0$                                                             & $0$                                            & $2.8\pm0.2$                 & $16.8^{+9.7}_{-8.8}$     & $14198\pm169$ & $104\pm14$    & $1.3\pm0.2$ \\[2pt]
	$2.6\pm0.4$ & on             & $0$                                                             & $0$                                            & $2.8\pm0.2$                & $14.8^{+6.6}_{-6.9}$      & $14202\pm169$ & $104\pm14$    & $1.3\pm0.2$ \\[2pt]
	$2.6\pm0.4$ & off              & $104.8^{+21.9}_{-35.6}$                       & $0$                                            & $2.8\pm0.2$                 & $17.1^{+9.1}_{-7.9}$     & $14185\pm168$ & $91\pm13$       & $1.1\pm0.2$ \\[2pt]
	$2.6\pm0.4$ & on              & $95.4\pm26.5$                                       & $0$                                            & $2.8\pm0.2$                 & $14.5^{+6.7}_{-7.2}$      & $14189\pm168$ & $91\pm13$       & $1.1\pm0.2$ \\[2pt]
	\hline \\
	\end{tabular}
\end{table*}

We found that allowing the eccentricity to float led to a negligible difference in $\chi^2_{red}$, and that the eccentricity values being found were within $2\sigma$ of 0. We therefore concluded that the small eccentricity values being returned were arising owing to the biases inherent in the MCMC method \citep{ford2006}, and that the orbit of WASP-25 is circular. In this we agree with \citetalias{enoch2011}. We confirmed this conclusion regarding a circular orbit using the F-test of \citet{lucy1971}, which returned very high probabilities of the small eccentricity values having arisen by chance.

We found little difference between the quality of fit for the equivalent cases with the MS mass-radius relation forced on the system, and those without the same constraint. The relaxation of this prior leads to larger values of $\lambda$, but also increases the discrepancy between the stellar mass and radius values. The stellar mass value varies little between runs, but relaxing the MS prior reduces the stellar radius by between $2$ and $3$\,percent, dependent on the other priors being applied. This leads to an increase in the stellar density of between $7$ and $12$ percent from $\bar{\rho}_{*,\,MS}\approx1.22\rho_\odot$ to $\bar{\rho}_{*,\,no\,MS}\approx1.34\rho_\odot$, averaged across all combinations of the other priors. Considering the impact parameter, we find that relaxing the MS requirement gives a value of $\bar{b}=0.38^{+0.16}_{-0.22}$, whilst using the prior returns $\bar{b}=0.44^{+0.11}_{-0.12}$, both averaged across all other combinations of priors. The S parameter increases from an average of $3.56$ to $5.92$ when the prior is removed. In light of these differences, we elect to apply the MS prior in our final analysis.

\begin{figure*}
	\subfloat{
		\includegraphics[width=0.48\textwidth]{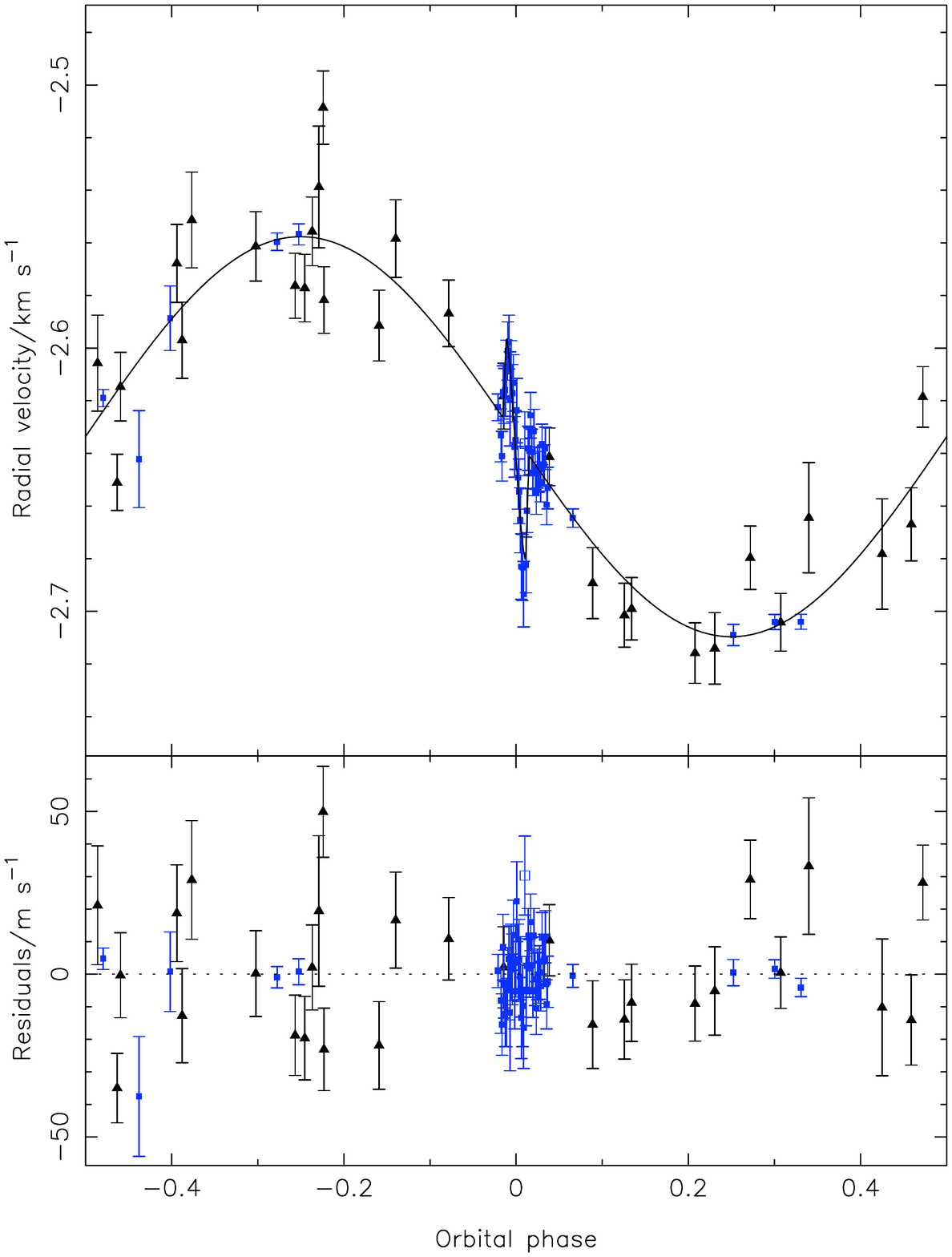}
		\label{fig:W25RV}}
	\subfloat{
		\includegraphics[width=0.48\textwidth]{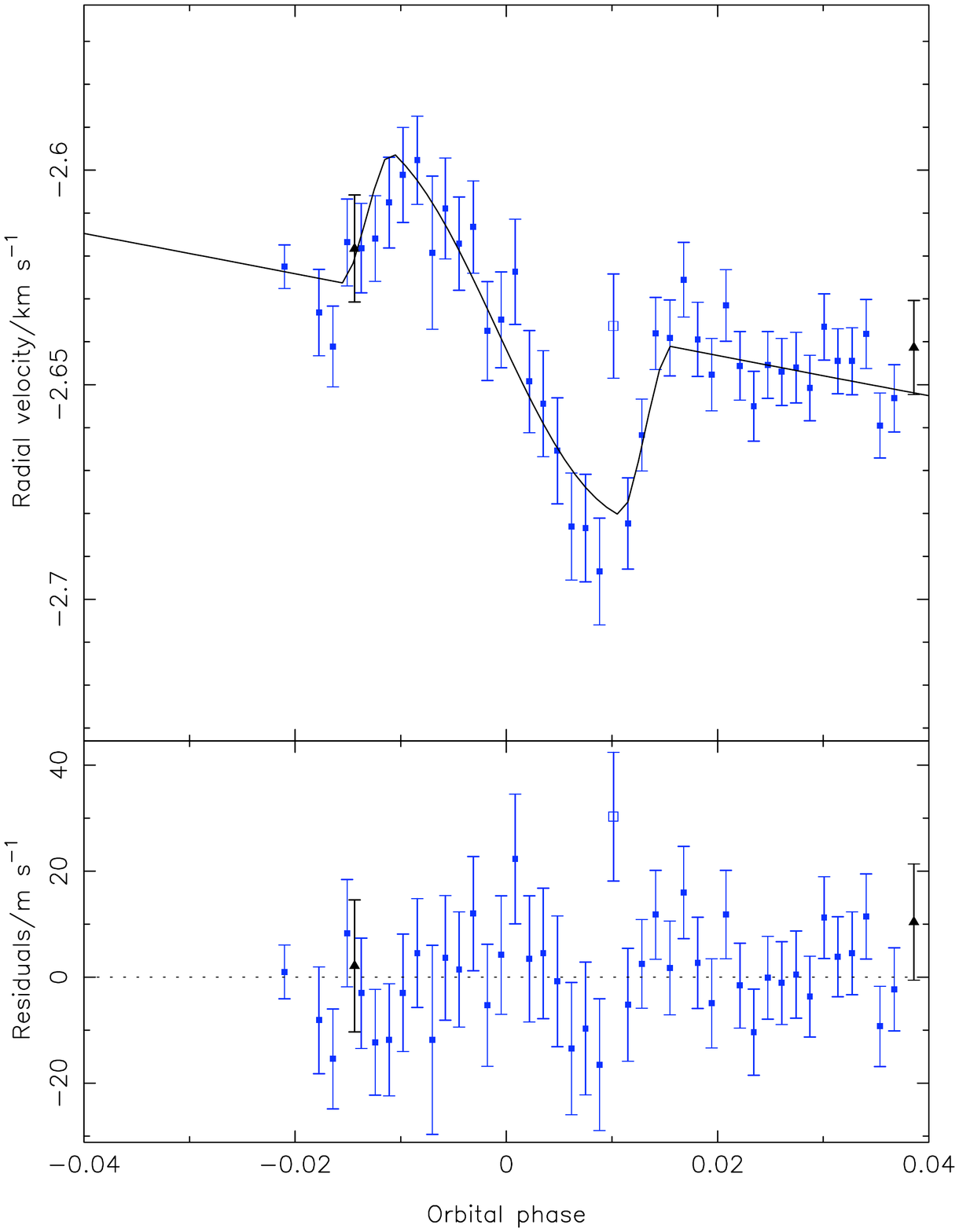}
		\label{fig:W25RM}}
	\caption{Results from the fit to the data for WASP-25 of our optimal solution: a circular orbit, no long-term RV trend and no prior on the spectroscopic $v\sin I$. The main sequence mass-radius relation was not enforced. The point denoted by the open square was found to lie $3\,\sigma$ from the best-fitting model, and was not included in the analysis. Legend as for Fig.\,\ref{fig:W16res}.}
	\label{fig:W25res}
\end{figure*}

Adding a long-term linear trend in RV improved the $\chi^2_{spec}$ of the solution, but the value of the trend varied significantly between runs, ranging from $\approx2$ to $\approx105$\,m\,s$^{-1}$\,yr$^{-1}$. We also found that in some cases the models produced when a trend was applied showed a notable offset from the RV data in transit. To check whether a trend was truly present in the system, 2 additional RV measurements were obtained using HARPS on 2010 August 25 and 26. Analysing these in conjunction with previously obtained data shows no evidence for a long-term RV trend, and so we disregard this possibility for our final solution. Introducing a prior on the spectroscopic $v\sin I$ produced no improvement to the quality of fit to the data, irrespective of the other flags. We do not therefore apply such a prior in our final solution, and take this opportunity to obtain a separate measurement of the projected stellar rotation speed.

\begin{figure*}
	\subfloat{
		\includegraphics[width=0.48\textwidth]{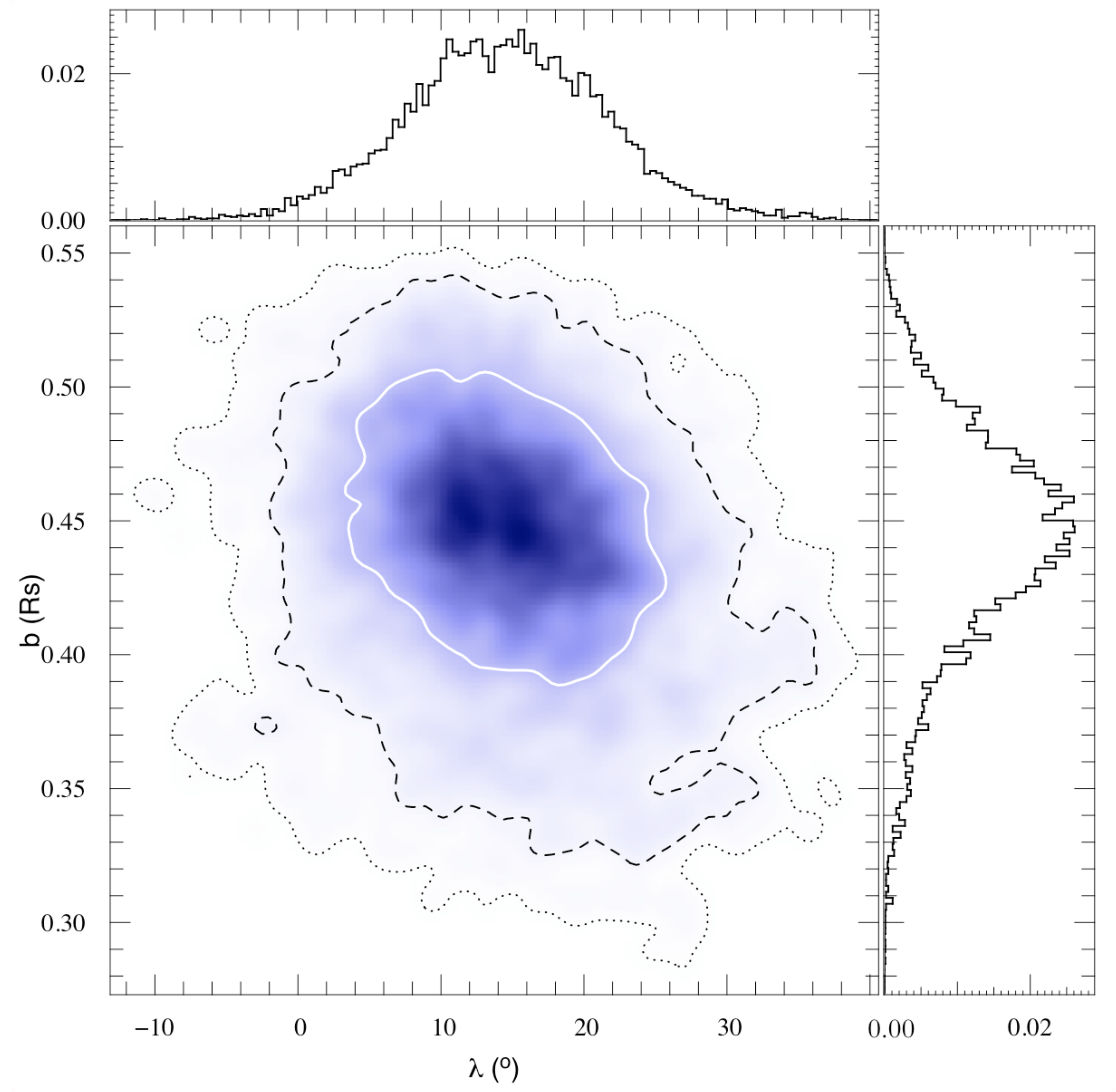}
		\label{fig:W25prob2}}
	\subfloat{
		\includegraphics[width=0.48\textwidth]{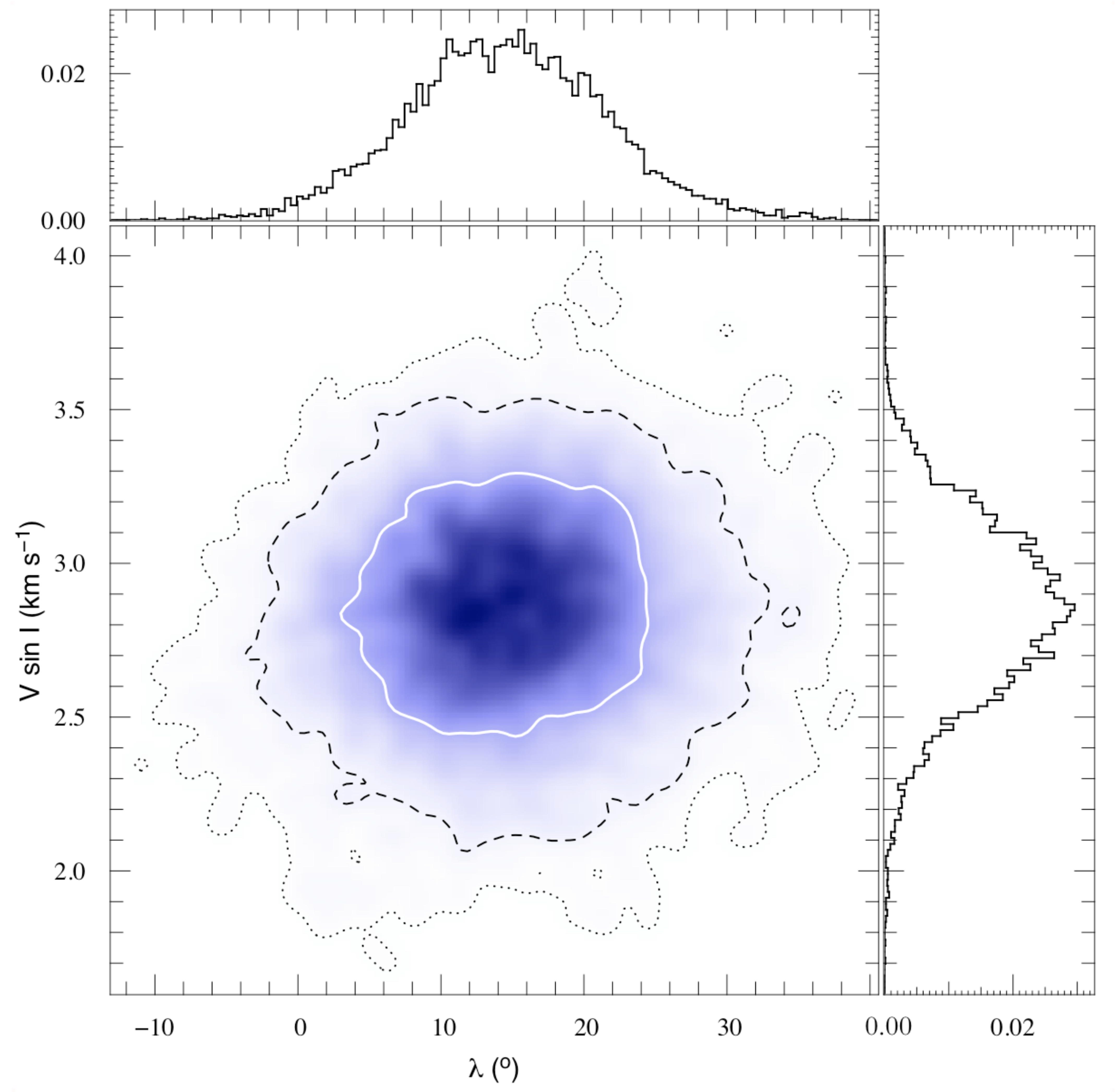}
		\label{fig:W25prob5}}
	\caption{Posterior probability distributions, derived from the Markov chain, for the fit to the data for WASP-25 described in Fig.\,\ref{fig:W25res}. Key as for Fig.\,\ref{fig:W16prob}. \textit{Left:} $b$ and $\lambda$. There are appears to be some small level of correlation between the two parameters. \textit{Right:} $v\sin I$ and $\lambda$. $\lambda=0$ falls outwith the $68.27$\,percent confidence contour, but within the $95.45$\,percent confidence contour, indicating a moderately significant detection of $\lambda$.}
	\label{fig:W25prob}
\end{figure*}

Taking the results of these investigations into account, we select the solution with $e=0$, no long-term linear trend in RV and no prior on $v\sin I$, with the main sequence mass-radius relation enforced. This gives $\lambda=14.6^{\circ}\pm6.7$, a detection of the RM effect at $2.2\sigma$ from $0$. We also obtain a value for the stellar rotation of $v\sin I=2.9\pm0.3$\,km\,s$^{-1}$, slightly greater than but in agreement with our updated spectroscopic value of $2.6\pm0.4$\,km\,s$^{-1}$. The impact parameter for this solution is $0.44\pm0.04$. No correlation is apparent between $v\sin I$ and $\lambda$ (see figure\,\ref{fig:W25prob5}), although there is evidence for a correlation between the impact parameter and $\lambda$ (see figure\,\ref{fig:W25prob2}). It is possible that this correlation is responsible for the poor fit of the model to some parts of the RM data.

The mechanism responsible for the outlier that we omitted from our analysis is unknown, although we note that \citet{simpson2011} experienced a similar situation in their analysis of the WASP-38 system, positing seeing changes of telescope guiding faults as possible causes. We suggest a third mechanism; the discrepant point might be caused by the planet traversing a stellar spot. In such a situation the spot would mask the presence of the planet, causing the RV measurement to diverge from the standard RM anomaly pattern. This scenario was suggested to explain a similar anomaly in the data for the WASP-3 system \citep{tripathi2010}, but we note that the divergence from the RM effect in that case showed a gradual rise and fall rather than the delta function change observed here, and was eventually attributed to the effect of moonlight. Unfortunately we lack simultaneous photometry from the night of the observed spectroscopic transit, which would show the presence of such a spot. It is also possible that some form of transient event, such as a white light stellar flare, is responsible for the drastic, sudden change in measured RV for this point, although. Such events were discussed in the context of LQ Hya \citep{montes1999}, and were observed to produce chromospheric disturbance in the core of otherwise normal spectral lines. This dilution of the spectral lines could affect the continuum level during the flare event, and potentially lead to anomalous redshifting for a short period of time. Such an event would have to be very short duration however, and coincide with the planet's path across the stellar disc.

Should we consider WASP-25 to be aligned? \citetalias{winn2010} put forward a criterion of $\lambda\geq10^o$ to $>3\sigma$ for misalignment; our result for WASP-25 clearly fails this test. \citetalias{triaud2010} suggest an alternative criterion of $\lambda>30^o$ as the limit above which we can be sure a system is misaligned given the average magnitude of the errors in $\lambda$ that are found by analysis of the RM effect. WASP-25 also misses this target by some margin. But the data for the RM effect appear to be slightly asymmetric in Fig.\,\ref{fig:W25RM}, suggesting that the system is misaligned (although we note that the best fitting model does not reflect this).

This slight asymmetry in the RM anomaly might arise as a result of some form of systematic effect. We have already mentioned the possibility of stellar spots in the context of the anomalous datum omitted from our analysis. Could they also provide a possible explanation for the asymmetry? Consider a star on which stellar spots are more numerous in one hemisphere than the other during the planetary transit, but on which they lie away from the transit chord. As the planet transits the more spotty hemisphere it will hide a comparatively larger fraction of the photosphere, and therefore mask a greater contribution to the overall flux than when it is transiting the less spotty hemisphere. The half of the anomaly corresponding to the spotted hemisphere would therefore have a greater amplitude than the half of the anomaly corresponding to the unspotted hemisphere, leading to an asymmetric RM effect. If the difference in the number and/or size of spots between the two hemispheres is small then the asymmetry would be only minor. This interesting systematic was discussed by \citet{albrecht2011} for the case of WASP-2, and also seems to have played a role in the analysis of the CoRoT-2 RM in \citet{bouchy2008}. In the case of WASP-25 the approaching, blue-shifted hemisphere would be required to have a slightly greater density of stellar spots than the receding, red-shifted hemisphere, which would also lead back to the possibility of a transient event being responsible for the anomalous datum.

We will return to the question of WASP-25's alignment in section\,\ref{sec:misalign}.

\subsection{WASP-31}
\label{sec:W31res}
WASP-31 \citep{anderson2011} is a bloated, $0.5$\,$M_J$ planet orbiting an F-type star of sub-solar metallicity with a period of $3.5$\,days. The host star is a moderately rapid rotator, with $v\sin I=8.1\pm0.5$ from spectroscopy. Full results of our analysis can be found in Table\,\ref{tab:W31res}.

\begin{table*}
	\caption{A comparison of the $\chi^2$ and $\chi^2_{red}$ values for WASP-31 for each combination of Bayesian priors. All values of $\chi^2$ include the Bayesian penalties applicable for that combination of priors.}
	\label{tab:W31res}
	\begin{tabular}{lllllllll}
	\hline \\
	$v\sin I$ prior & MS prior & $\dot{\gamma}$/ms$^{-1}$yr$^{-1}$ & eccentricity                              & $v\sin I$/km\,s$^{-1}$ & $\lambda$/$^{\circ}$ & $\chi^2$                & $\chi^2_{RV}$ & $\chi^2_{red}$ \\
	\hline \\
	off                     & off            & $0$                                                           & $0.027^{+0.032}_{-0.019}$ & $7.5\pm0.8$                 & $2.8^{+1.1}_{-2.9}$   & $14703\pm171$ & $64\pm11$       & $0.9\pm0.2$ \\[2pt]
	off                     & on            & $0$                                                          & $0.031^{+0.029}_{-0.019}$ & $7.7^{+0.9}_{-0.8}$     & $3.6^{+2.9}_{-3.5}$   & $14708\pm172$ & $64\pm11$       & $0.9\pm0.2$ \\[2pt]
	off                     & off            & $6.1^{+8.1}_{-8.4}$                               & $0.023^{+0.031}_{-0.017}$ & $7.4\pm0.7$                 & $2.8^{+2.9}_{-2.8}$   & $14700\pm171$ & $63\pm11$       & $0.9\pm0.2$ \\[2pt]
	off                     & on            & $12.6^{+8.4}_{-7.6}$                            & $0.037^{+0.035}_{-0.016}$  & $7.8\pm0.8$                 & $3.1^{+3.0}_{-2.8}$   & $14695\pm171$ & $63\pm11$       & $0.9\pm0.2$ \\[2pt]
	off                     & off            & $0$                                                           & $0$                                           & $7.5\pm0.7$                 & $2.7\pm3.0$                & $14702\pm171$ & $64\pm11$        & $0.9\pm0.2$ \\[2pt]
	off                     & on            & $0$                                                          & $0$                                            & $7.5\pm0.7$                 & $2.8\pm3.1$                & $14706\pm172$ & $64\pm11$       & $0.9\pm0.2$ \\[2pt]
	off                     & off            & $6.4^{+7.9}_{-8.1}$                               & $0$                                            & $7.5\pm0.7$                 & $2.4^{+2.9}_{-2.7}$   & $14698\pm171$ & $63\pm11$        & $0.9\pm0.2$ \\[2pt]
	off                     & on            & $5.3^{+8.8}_{-7.7}$                              & $0$                                            & $7.3^{+0.7}_{-0.6}$     & $3.0^{+3.4}_{-3.1}$   & $14698\pm171$ & $64\pm11$        & $0.9\pm0.2$ \\[2pt]
	$8.1\pm0.5$ & off             & $0$                                                           & $0.023^{+0.029}_{-0.017}$  & $7.9\pm0.4$                  & $2.5^{+2.8}_{-2.6}$   & $14693\pm171$ & $64\pm11$       & $0.9\pm0.2$ \\[2pt]
	$8.1\pm0.5$ & on             & $0$                                                           & $0.041^{+0.033}_{-0.027}$  & $8.0\pm0.5$                  & $3.2^{+3.0}_{-2.9}$  & $14703\pm171$ & $64\pm11$        & $0.9\pm0.2$ \\[2pt]
	$8.1\pm0.5$ & off             & $-0.1^{+9.2}_{-6.6}$                              & $0.022^{+0.033}_{-0.016}$   & $7.9^{+0.4}_{-0.5}$    & $2.7^{+2.9}_{-2.7}$   & $14698\pm171$ & $63\pm11$        & $0.9\pm0.2$ \\[2pt]
	$8.1\pm0.5$ & on             & $3.4^{+5.7}_{-4.6}$                               & $0.038^{+0.023}_{-0.018}$   & $8.0\pm0.4$                 & $3.0\pm2.7$               & $14702\pm171$ & $64\pm11$        & $0.9\pm0.2$ \\[2pt]
	$8.1\pm0.5$ & off             & $0$                                                           & $0$                                             & $7.9\pm0.4$                  & $2.8^{+2.7}_{-2.9}$  & $14697\pm171$ & $64\pm11$        & $0.9\pm0.2$ \\[2pt]
	$8.1\pm0.5$ & on             & $0$                                                           & $0$                                             & $7.8\pm0.4$                  & $3.0^{+3.0}_{-2.9}$  & $14701\pm171$ & $65\pm11$        & $0.9\pm0.2$ \\[2pt]
	$8.1\pm0.5$ & off             & $6.1^{+10.3}_{-8.6}$                             & $0$                                             & $7.8\pm0.4$                   & $2.7^{+2.7}_{-2.9}$ & $14701\pm171$ & $64\pm11$        & $0.9\pm0.2$ \\[2pt]
	$8.1\pm0.5$ & on             & $5.4^{+7.7}_{-8.5}$                               & $0$                                             & $7.9\pm0.4$                   & $3.0^{+3.0}_{-2.9}$ & $14705\pm171$ & $65\pm11$        & $0.9\pm0.2$ \\[2pt]
	\hline \\
	\end{tabular}
\end{table*}

We found no difference between the $\chi^2_{red}$ values for any combination of priors. We found that imposing the main-sequence mass-radius relation had little effect on the fit to the spectroscopic data, but had a deleterious effect on the stellar parameters. Removing the prior produced an increase in stellar radius of between $3$ and $6$\,percent and a decrease in the stellar mass of between $1$ and $2$\,percent, leading to a decrease in stellar density of between $8$ and $15$\,percent from $\bar{\rho}_{*,\,MS}\approx0.67\rho_\odot$ to $\bar{\rho}_{*,\,no\,MS}\approx0.62\rho_\odot$, averaged across all other combinations of priors. Comparing the impact parameter and $S$ statistic, we find $\bar{b}=0.79^{+0.03}_{-0.05}$ and $\bar{S}=10.2$ with no MS prior applied, and $\bar{b}=0.77^{+0.03}_{-0.04}$ with $\bar{S}=2.9$ when the requirement for the star to be on the MS is enforced. Owing to the much more favourable $S$\,statistic, and the influence on the stellar parameters, we elect to use results which account for the MS relationship. Adding a linear velocity trend gave no discernible difference in the quality of the fit to the spectroscopic data, and with a magnitude of $|\dot{\gamma}|<13$\,m\,s$^{-1}$\,yr$^{-1}$ we conclude that no such trend is present in the system. Adding a prior on the spectroscopic $v\sin I$ made little difference to the results despite the relatively rapid rotation, so we again choose the simpler route and neglect such a prior. Finally, we choose a circular solution; the F-test of \citet{lucy1971} shows that the small eccentricity values returned when $e$ is allowed to float are insignificant.

\begin{figure*}
	\subfloat{
		\includegraphics[width=0.48\textwidth]{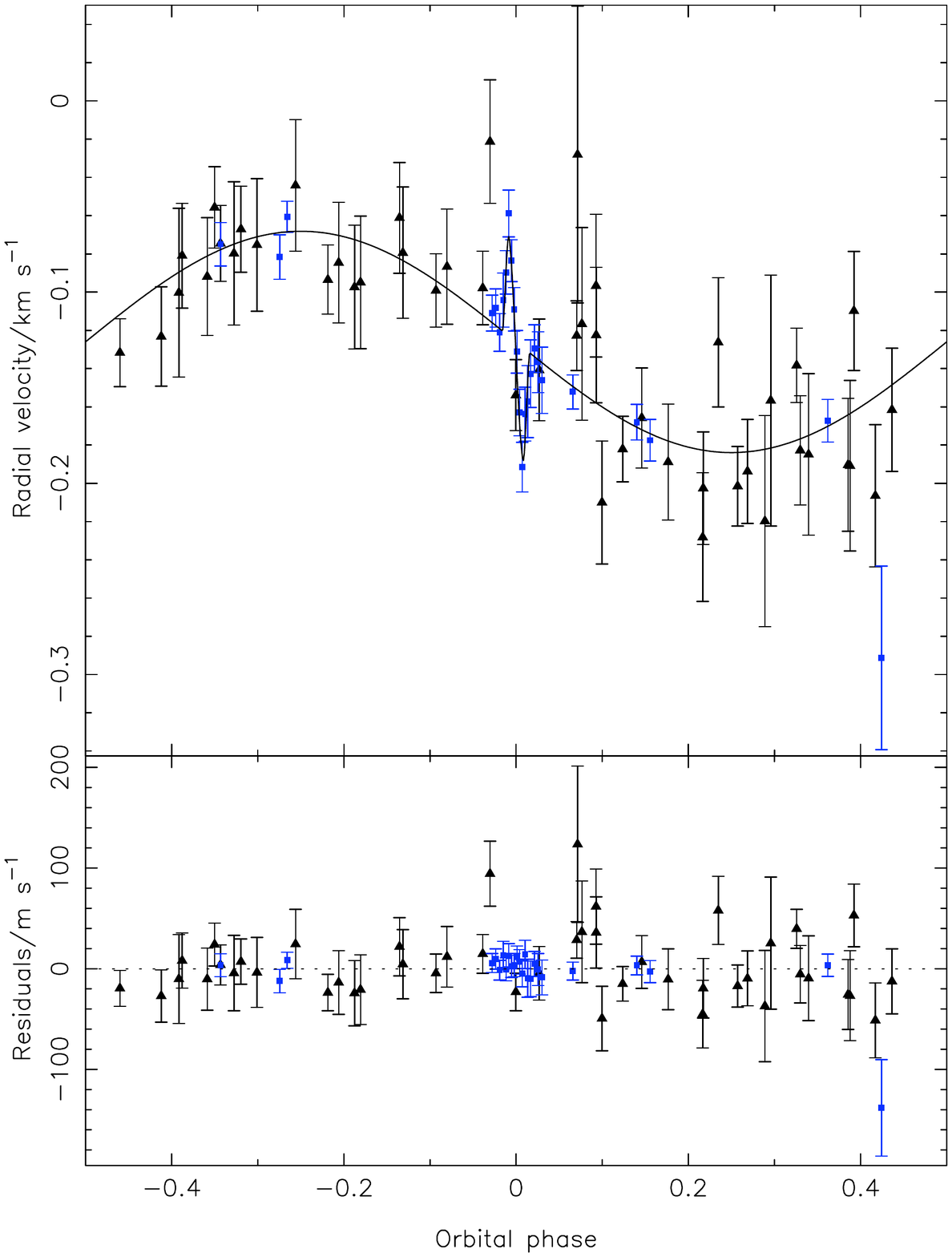}
		\label{fig:W31RV}}
	\subfloat{
		\includegraphics[width=0.48\textwidth]{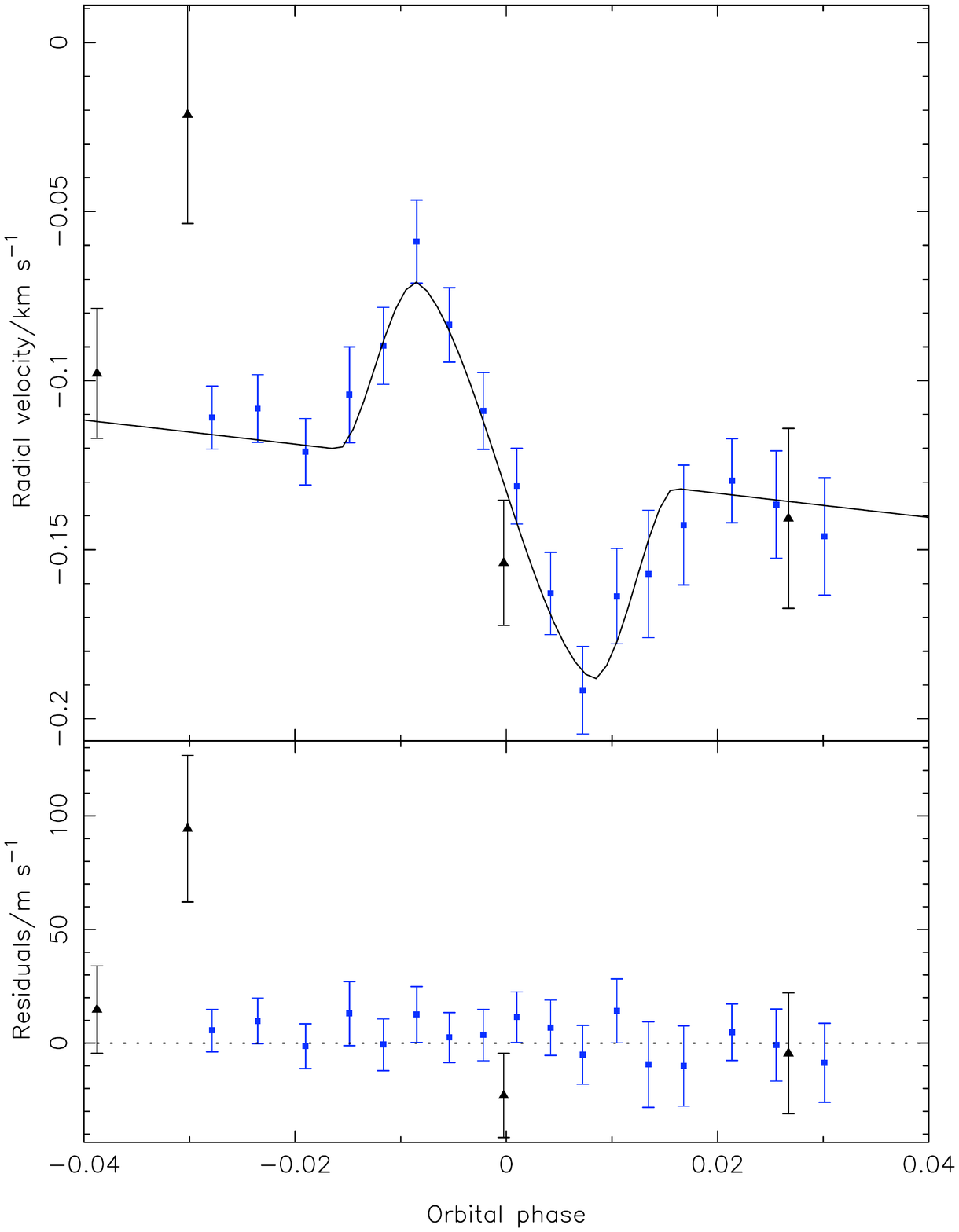}
		\label{fig:W31RM}}
	\caption{Results from the fit to the data for our adopted solution for WASP-31, with a circular orbit, no prior on the spectroscopic $v\sin I$, no long-term radial velocity trend, and the mass-radius relationship applied. Legend as for Fig.\,\ref{fig:W16res}.}
	\label{fig:W31res}
\end{figure*}

\begin{figure*}
	\subfloat{
		\includegraphics[width=0.48\textwidth]{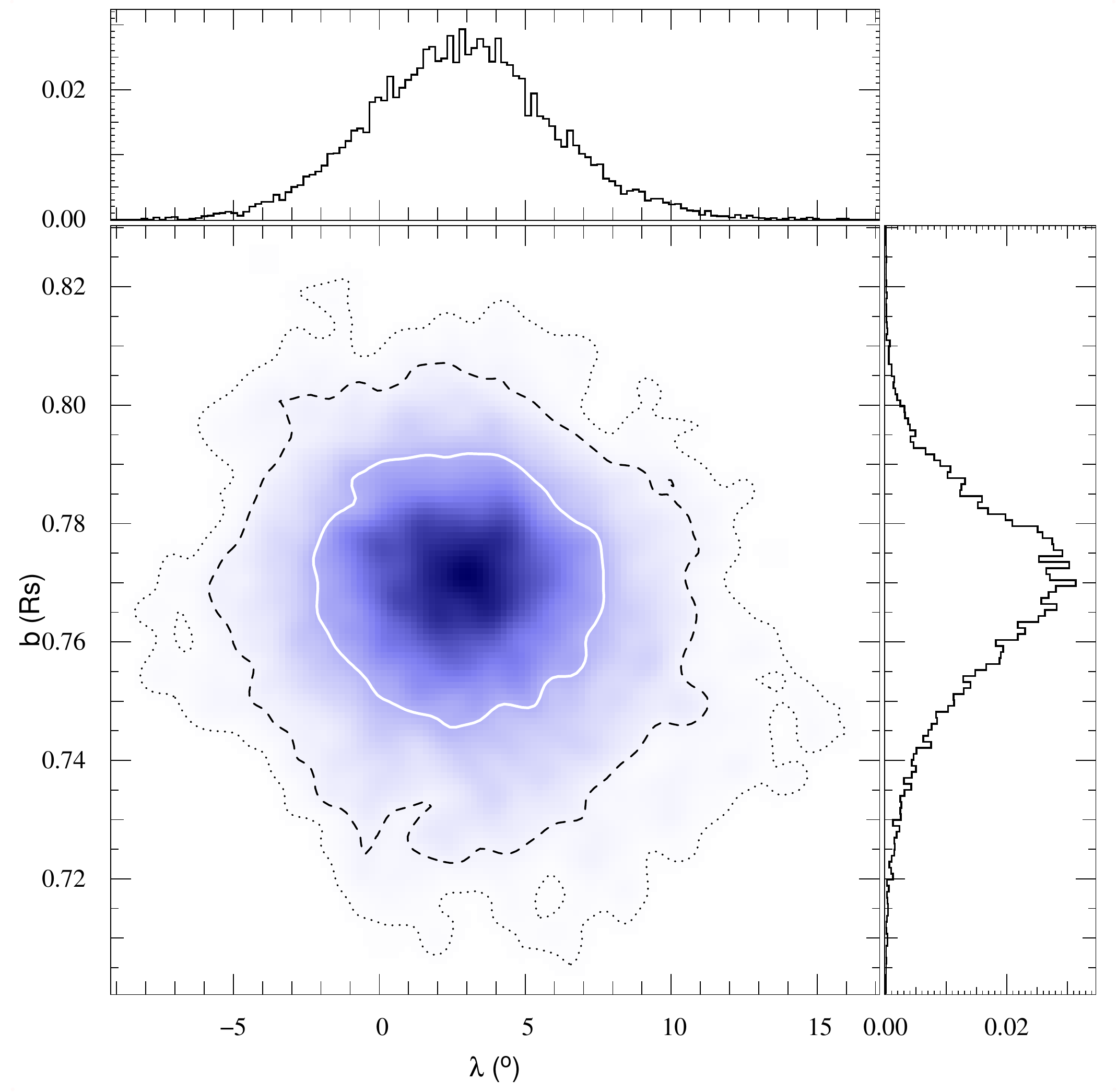}
		\label{fig:W31prob2}}
	\subfloat{
		\includegraphics[width=0.48\textwidth]{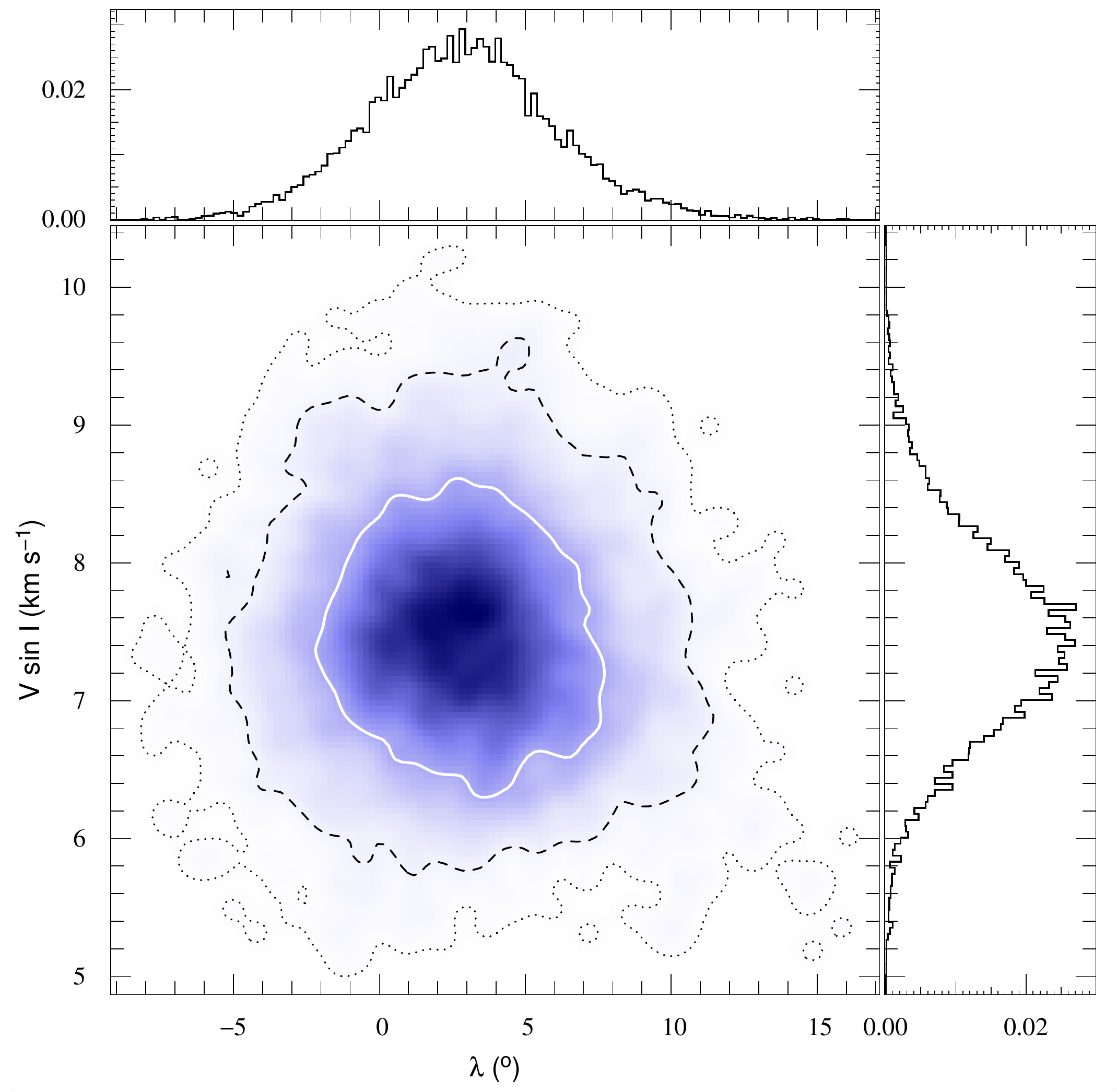}
		\label{fig:W31prob5}}
	\caption{Posterior probability distributions, derived from the Markov chain, for the fit to the data for WASP-31 described in Fig.\,\ref{fig:W31res}. Key as for Fig.\,'ref{fig:W16prob}. \textit{Left:} $b$ and $\lambda$. \textit{Right:} $v\sin I$ and $\lambda$. $\lambda=0$ lies well within the main body of the distribution.}
	\label{fig:W31prob}
\end{figure*}

Our optimal solution is therefore that obtained with no $v\sin I$ prior, no velocity trend, the MS prior active, and $e=0$. This set of priors gives $\lambda=2.8^{\circ}\pm3.1$ , leading to the conclusion that the WASP-31 system is well-aligned. It is worth noting that this would be the conclusion whichever combination of priors we chose, as all of the values of $\lambda$ that we obtained lie within $1.2\sigma$ of $0^o$. The impact parameter is $0.77^{+0.01}_{-0.02}$. The stellar rotation for this solution has a value of $v\sin I=7.5\pm0.7$\,km\,s$^{-1}$. As with our result for WASP-16 this is slower than the spectroscopic value, but in this case the value agrees to within $1\sigma$. Again, an alternative analysis using the calibration of \citet{gray1992} returns a value of $v\sin I$ ($7.5\pm0.5$\,km\,s$^{-1}$) more similar to our MCMC result.
WASP-31 is not included in the sample of \citet{schlaufman2010} owing to its time of publication. In order to check the possibility of misalignment along the line-of-sight, we follow the method of \citeauthor{schlaufman2010} and calculate the rotation statistic, $\Theta$. The age of WASP-31\,A is somewhat uncertain however; its lithium abundance, gyrochronology and the presence of a close companion all suggest ages of $\approx1$\,Gyr, whilst previous stellar model fits imply an older age of $4\pm1$\,Gyr. We reassess the isochronal fit for the system, obtaining ages of $4.0^{+1.8}_{-1.0}$\,Gyr (Padova models), $2.8^{+1.4}_{-1.0}$\,Gyr (Yonsei-Yale models), $3.5^{+2.3}_{-1.3}$\,Gyr (Teramo models) and $2.8^{+1.6}_{-1.2}$\,Gyr (VRSS models). Using these estimates we calculate values of $\Theta=-4.5$, $-3.0$, $-3.7$ and $-2.8$ respectively; WASP-31 is therefore rotating more rapidly than expected given its age in both cases. The chance of significant misalignment along the line-of-sight therefore seems slim; the inclination of the WASP-31\,b's orbit is $84.6\pm0.2^{\circ}$, leaving little room for an increase in rotation velocity owing to line-of-sight misalignment.

\section{Integration into the ensemble of results}
\label{sec:discuss}
\begin{figure}
	\includegraphics[width=0.50\textwidth]{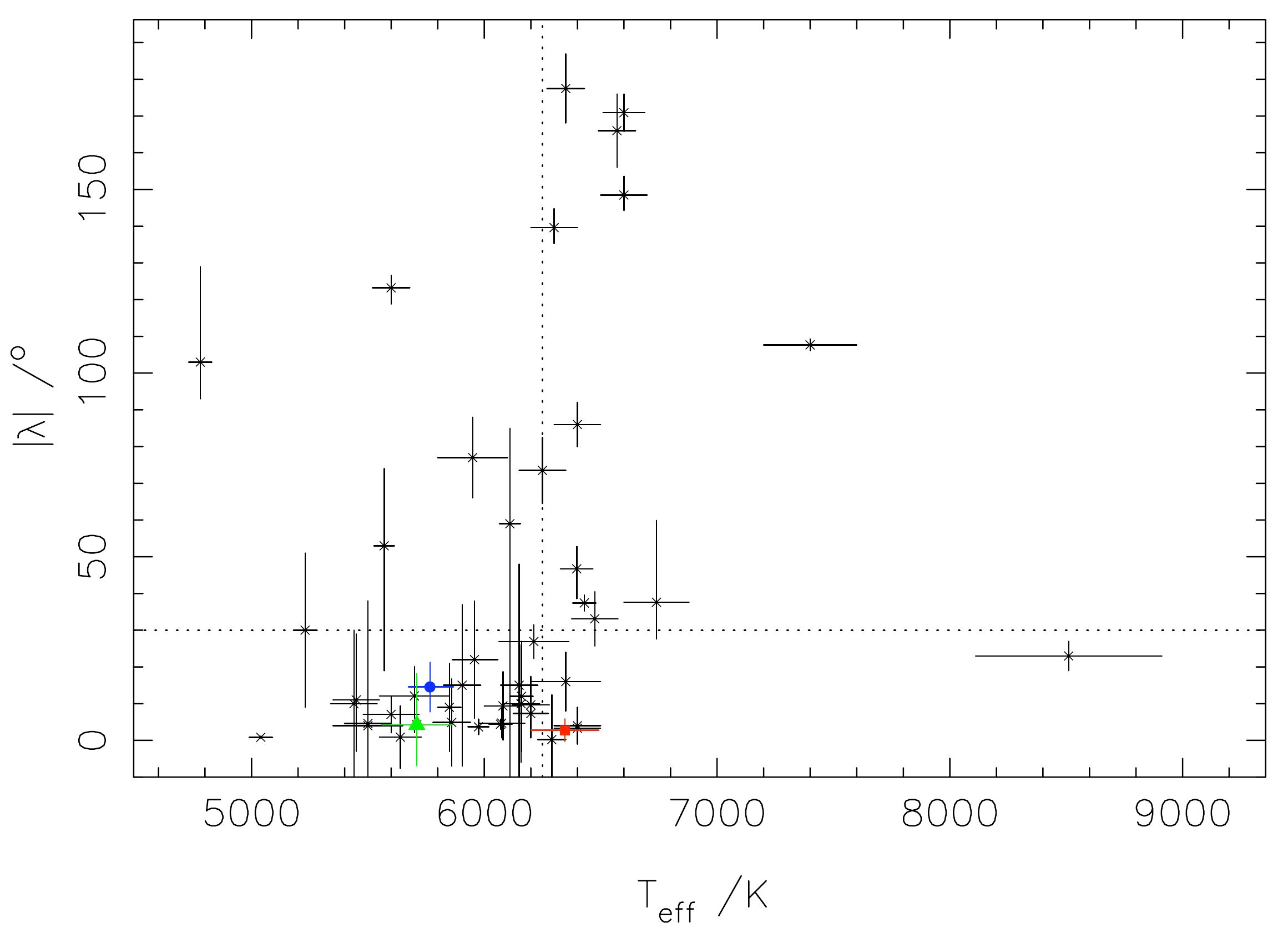}
	\caption{Projected stellar obliquity, $\lambda$, as a function of stellar effective temperature for all systems with confirmed measurements. WASP-16 is marked by a green, filled triangle,  WASP-25 by a blue, filled circle, and WASP-31 by a red, filled square. The vertical dotted line marks the distinction between `cool' and `hot' systems, whilst the horizontal dotted line marks $30^{\circ}$, the angle above which a system is considered to be misaligned in \citetalias{winn2010} and \citetalias{triaud2010}.}
	\label{fig:angleT}
\end{figure}

The analysis of \citetalias{winn2010} provides a good starting point for integrating our new results into the existing ensemble of RM measurements. Fig.\,\ref{fig:angleT} reproduces fig.\,2 from their paper, with the addition of all complete RM measurements made since its publication (except WASP-23 \citep{triaud2011}, for which the result is still highly uncertain, and WASP-26 \citep{anderson2011b}, which showed only a very low amplitude and was classed as a non-detection); we list these planets in Table\,\ref{tab:newRM}. We also elect to include most of the systems that \citetalias{winn2010} disregard during their analysis as having insufficiently precise measurements of $\lambda$ \footnote{HAT-P-2, CoRoT-1, CoRoT-3, HD149026, Kepler-8, TrES-1 and TrES-2. See references within \citetalias{winn2010}. Although WASP-2 has a measured value for $\lambda$, the most recent analysis of the system failed to detect a signal \citep{albrecht2011} and thus we continue to exclude this system.} in order to provide a full picture of the current state of RM analysis. Whilst it is true that making a definitive statement regarding alignment is more difficult for these systems owing to their large uncertainties, the criteria for granting misaligned status should take account of this. We are also interested in comparing our new measurements to the general form of the current ensemble. Omitting the systems listed above does not simplify this task, so we elect to include them.

\begin{table*}
	\caption{Relevant data for the planetary systems for which the Rossiter-McLaughlin effect has been characterised since the publication of \citetalias{winn2010}. We add these systems to the \citeauthor{winn2010} sample to bring the ensemble of results up to date and allow us to better analyse the place of WASP-25 and WASP-31 within that ensemble.}
	\label{tab:newRM}
	\begin{tabular}{llllll}
		\hline \\
		System   & $i$/$^{\circ}$       & $v\sin I$/km\,s$^{-1}$ & $T{\rm eff}$/K & $\lambda$/$^{\circ}$ & Reference \\
		\hline \\
		CoRoT-18 & $86.5^{+1.4}_{-0.9}$ & $8.0\pm1.0$ & $5440\pm100$ & $10\pm20$ & \citet{hebrard2011b} \\[2pt]
		HAT-P-4  & $88.76^{+0.89}_{-1.38}$ & $5.83\pm0.35$          & $5860\pm80$    & $4.9\pm11.9$            & \citet{winn2011} \\[2pt]
		HAT-P-6  & $85.51\pm0.35$          & $7.5\pm1.6$            & $6570\pm80$    & $166\pm10$              & \citet{hebrard2011} \\[2pt]
		HAT-P-8  & $87.5\pm1.4$            & $14.5\pm0.8$           & $6200\pm80$    & $-17^{+9.2}_{-11.5}$    & \citet{latham2009,moutou2011} \\[2pt]
		HAT-P-9  & $86.5\pm0.2$            & $12.5\pm1.8$           & $6350\pm150$   & $-16\pm8$               & \citet{shporer2009,moutou2011} \\[2pt]
		HAT-P-11 & $89.17^{+0.46}_{-0.60}$ & $1.00^{+0.95}_{-0.56}$ & $4780\pm50$    & $103^{+26}_{-10}$       & \citet{winn2010b} \\[2pt]
		HAT-P-14 & $83.52\pm0.22$          & $8.18\pm0.49$          & $6600\pm90$    & $189.1\pm5.1$           & \citet{winn2011} \\[2pt]
		HAT-P-16 & $86.6\pm0.7$            & $3.9\pm0.8$            & $6158\pm80$    & $-10.0\pm16$            & \citet{buchhave2010,moutou2011} \\[2pt]
		HAT-P-23 & $85.1\pm1.5$            & $7.8\pm1.6$            & $5905\pm80$    & $15\pm22$               & \citet{bakos2011,moutou2011} \\[2pt]
		HAT-P-30 & $83.6\pm0.4$            & $3.07\pm0.24$          & $6304\pm88$    & $73.5\pm9.0$            & \citet{johnson2011} \\[2pt]
		KOI-13.01 & $85.0\pm0.4$           & $65\pm10$                & $8511\pm400$       & $23\pm4$                   & \citet{barnes2011} \\[2pt]
		WASP-1   & $90\pm2$                & $0.7^{+1.4}_{-0.5}$    & $6110\pm45$    & $-59^{+99}_{-26}$       & \citet{albrecht2011} \\[2pt]
		WASP-7 & $87.2^{+0.9}_{-1.2}$ & $14\pm2$                   & $6400\pm100$   &  $86\pm6$                     & \citet{southworth2011,albrecht2012} \\[2pt]
		WASP-19  & $79,4\pm0.4$            & $4.63\pm0.26$          & $5500\pm100$   & $4.6\pm5.2$             & \citet{hellier2011} \\[2pt]
		WASP-22  & $88.26\pm0.91$          & $4.42\pm0.34$          & $5958\pm98$    & $22\pm16$               & \citet{anderson2011b} \\[2pt]
		WASP-24  & $83.64\pm0.29$          & $7.0\pm0.9$            & $6075\pm100$   & $-4.7\pm4.0$            & \citet{simpson2011} \\[2pt]
		WASP-38  & $88.83^{+0.51}_{-0.55}$ & $8.58\pm0.39$          & $6150\pm80$    & $15^{+33}_{43}$         & \citet{simpson2011} \\[2pt]
		XO-3     & $82.5\pm1.5$            & $18.4\pm0.2$          & $6430\pm50$    & $37.4\pm2.2$            & \citet{winn2009,hirano2011} \\[2pt]
		XO-4     & $88.8\pm0.6$            & $8.9\pm0.5$            & $6397\pm70$    & $-46.7\pm7.1$           & \citet{narita2010} \\[2pt]
		\hline \\
	\end{tabular}
\end{table*}

WASP-31 has an effective temperature of $6300\pm100$\,K, which falls with $1\sigma$ of the border between the `hot' and `cool' categories of \citetalias{winn2010}, albeit tending towards the `hot' side. We cannot therefore draw any conclusions as to how it affects the trend proposed in that paper.

With an effective temperature of $5750\pm100$\,K, WASP-25 falls into the `cool' category ($T{\rm eff}\leq6250$\,K) of \citetalias{winn2010}, which they find to be preferentially aligned -- their sample gives a probability of misalignment for `cool' stars of $0.17$. Updating this result using our expanded sample changes the probability to $0.20$ using either the criterion of \citetalias{winn2010}, or to $0.13$ using the criterion of $|\lambda|>30^o$ from \citetalias{triaud2010}. It is worth noting here that the apparently large differences in misalignment probability between the two criteria are an artefact of the sample size, which is still relatively small at 48 systems (30 `cool', 18 `hot'). Switching between the two criteria only changes the number of aligned systems by two for the `cool' sub-sample, and has no effect on the number of misaligned systems in the `hot' sub-sample. Under both criteria the apparent alignment of WASP-25\,b's orbit is in accordance with the \citetalias{winn2010} hypothesis. WASP-16, $T_{eff}=5700\pm150$ is also classified as a `cool' system. All available information points towards this system being well-aligned, and it therefore fits well with the hypothesis of \citetalias{winn2010}.

The final interesting point about Fig.\,\ref{fig:angleT} is the apparent lack of systems with mildly retrograde, close to polar orbits. There are currently only two systems with $80^o\leq\lambda\leq110^0$,  and only one more with $110^{\circ}\leq\lambda\leq140^{\circ}$ This relatively unpopulated region is less noticeable when considering $\psi$ owing to the increased size of the error bars, but it is still apparent. We speculate that truly polar orbits are perhaps unstable for some reason. Or perhaps it is simply our inability to determine the inclination of the stellar rotation axis that is at fault. It may be that some `aligned' systems actually have close to polar orbits if this angle is accounted for. It may also be that we simply have yet to observe very many systems in this region of the parameter space, and future publications may provide the data to fill this underpopulated area.

\begin{figure}
	\includegraphics[width=0.48\textwidth]{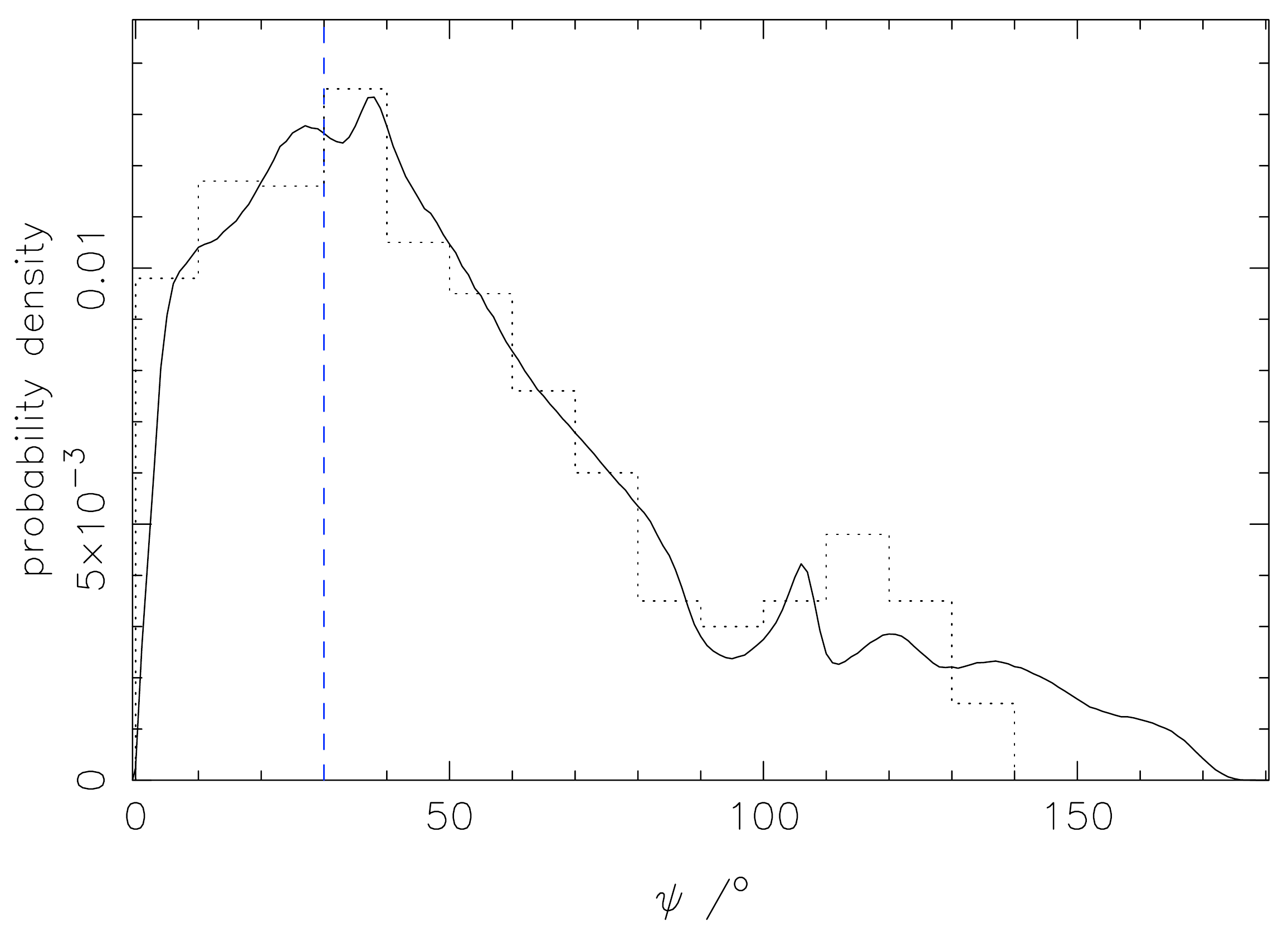}
	\caption{The total distribution of the true obliquity angle, $\psi$, for the complete sample of systems for which $\lambda$ has been measured. The dotted histogram represents the theoretical distribution of \citet{fabrycky2007}. The dashed line represents the limit of $\psi=30^{\circ}$ above which a system is considered to be misaligned. The overall forms are comparable, and the total $\psi$ distribution is similar to fig.\,10 of \citetalias{triaud2010}. The shape of the primary peak agrees well with theoretical predictions. The overall shape of the secondary, high angle peak in the distribution is less clear, but may be more shallow and broader than anticipated. The sudden drop in probability density at mid-range angles, around $\psi\approx90^o$, has become more pronounced when compared to the distribution of \citetalias{triaud2010}.}
	\label{fig:psitot}
\end{figure}

It has not been remarked upon before in this context, but a drop in the number of systems at mid-range obliquity angles is clearly predicted by the theoretical $\psi$ angular distribution of \citet{fabrycky2007}. It also clearly shows up in the angular distribution for the complete set of known obliquity angles, fig.\,10 in \citetalias{triaud2010}. We reproduce this figure in Fig.\,\ref{fig:psitot}, adding the probability distributions of the planets in Table\,\ref{tab:newRM} as well as those of the planetary systems from this study. $\psi$, the true misalignment angle, is given by
\begin{equation}
	\cos\psi = \cos I\cos i + \sin I\sin i\cos\lambda,
\label{eq:psi}
\end{equation}where $I$ is the inclination of the stellar rotation axis to the line-of-sight, and $i$ is the inclination of the orbital axis to the line-of-sight. To calculate the $\psi$ distribution for each planet we carried out $10^6$ Monte Carlo simulations, drawing values for $I$ from a uniform $\cos I$ distribution to represent the case in which stellar rotation axes are randomly oriented on the sky. We also accounted for the error bars on $i$ and $\lambda$ by drawing values from a Gaussian distribution with our optimal solution values as the mean values, and scaled to the uncertainties in those values. The individual planets' distributions were then summed to produce our total distribution, which is similar to that of \citetalias{triaud2010}, and still compares favourably to the theoretical histogram from \citet{fabrycky2007}. The drop in probability at mid-range angles is in line with the underpopulated region of Fig.\,\ref{fig:angleT}, and our additions bring the primary, low-angle peak closer in shape to the theoretical distribution. The overall shape of the secondary peak is less clear; it is still dominated by contributions from individual systems owing to the smaller number of planets with strongly misaligned orbits as compared to the number of aligned or weakly misaligned systems, but appears as though it may be broader and more shallow than the theoretical prediction.

Fig.\,\ref{fig:psitot} requires the assumption that the $I$, the stellar inclination, is isotropic and that the angular distribution is unimodal. However the discussion of \citetalias{winn2010} implies that the distribution is in fact bimodal. A clearer demonstration of the agreement between theoretical predictions and current observations is therefore to look at the distribution in $\lambda$. This requires the converse transformation of the predicted $\psi$ distribution of \citet{fabrycky2007} into $\lambda$.

We reproduce the lower panel of fig.\,9 from \citetalias{triaud2010}, taking into account the additional measurements of $\lambda$ from Table\,\ref{tab:newRM}. For HAT-P-7 and HAT-P-14, both of which have published $\lambda>180.0^{\circ}$, we used the negative angle equivalent ($360-\lambda$). This cumulative $\lambda$ distribution avoids both of the assumptions inherent in Fig.\,\ref{fig:psitot}. Agreement between the observational data and the theoretical predictions of \citet{fabrycky2007} has been improved, particularly for low- to mid-range angles, but the observational data are still slightly lacking in high obliquity systems compared to the theoretical histogram, whilst showing more low-obliquity systems than expected.

\begin{figure}
	\includegraphics[width=0.48\textwidth]{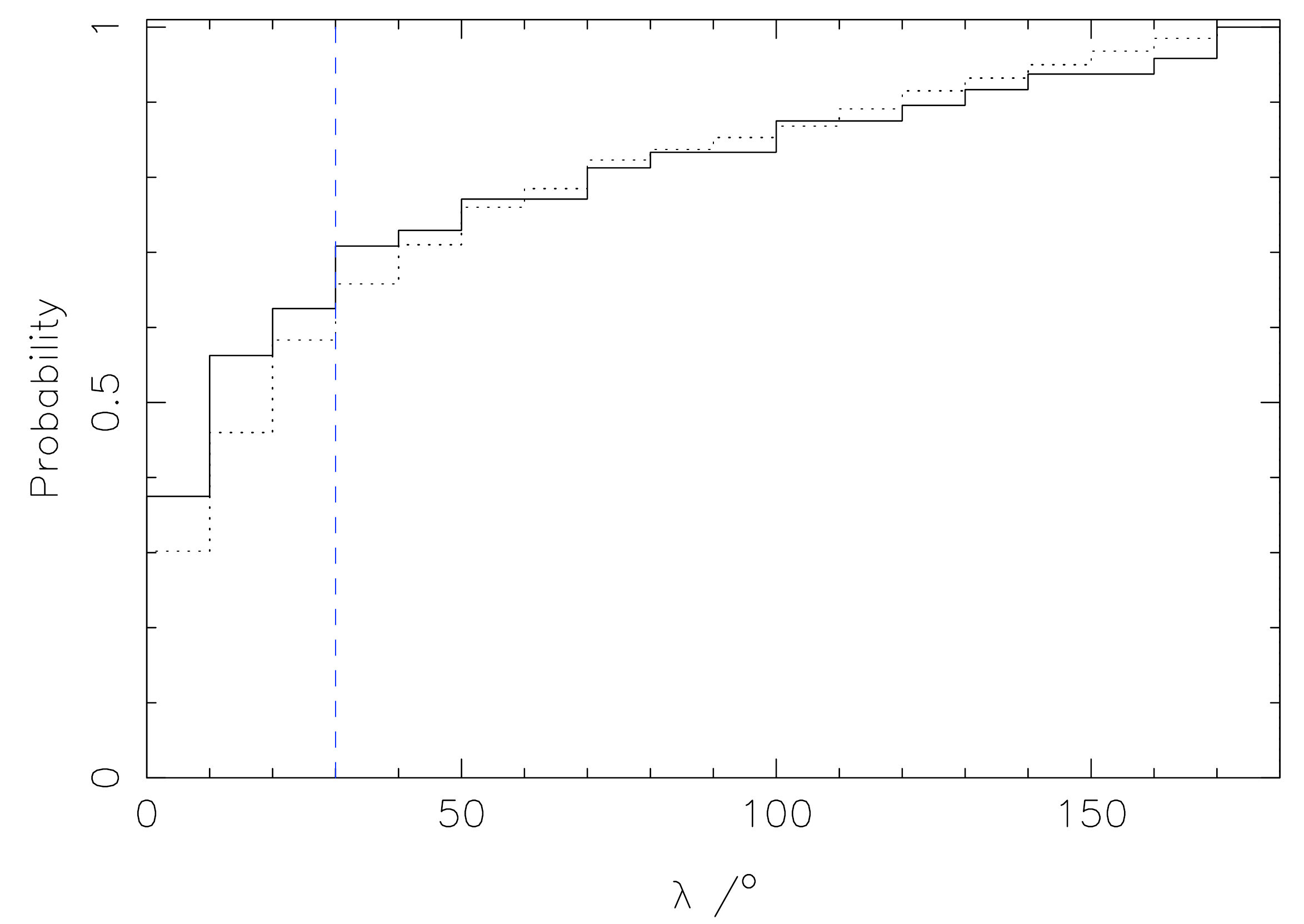}
	\caption{Cumulative probability histogram for $\lambda$. The solid line denotes observational data, whilst the dotted line denotes the theoretical distribution of \citet{fabrycky2007},  converted from $\psi$ to $\lambda$. The vertical blue, dashed line marks $\lambda=30^{\circ}$, the limit above which a planetary orbit is considered to be misaligned. The agreement between the two distributions has improved with the addition of measurements made since the publication of \citetalias{triaud2010}, particularly at mid-range angle, prograde orbits, but the observational data is still lacking in high obliquity systems compared to the theoretical prediction.}
	\label{fig:FThist}
\end{figure}

\section{A new misalignment test}
\label{sec:misalign}
The Rossiter-McLaughlin effect has now been measured for 48 transiting exoplanets, but as of yet there seems little consensus as to the best way of classifying them as aligned or misaligned. For most of the systems with measurements of $\lambda$ this is not a serious problem; either $|\lambda|>90^o$, or the error bars are such that the obliquity is consistent with zero. But as the number of RM measurements continues to grow, there will be an increasing number of systems in a similar situation to WASP-25, which exhibits a mildly asymmetrical RM anomaly but does not fulfil any of the current misalignment criteria.

There are two main criteria currently in use by the community. \citetalias{winn2010} use $|\lambda|>0^o$ at $\geq3\sigma$ significance to define a misaligned system. \citetalias{triaud2010} take $|\lambda|>30^o$ as their threshold, on the basis that errors in the obliquity angle are of the order of $10^o$, and therefore this gives $3\sigma$ significance as well. We would like to introduce a new test for misalignment that takes a completely different approach to these.

We consider the set of WASP planets for which the RM effect has been characterised using RV data, including the systems presented in this study. We neglect the WASP-33 system for which the misalignment angle has been measured only through Doppler tomography \citep{cameron2010b}, and disregard the ambiguous results for WASP-23 \citep{triaud2011} and WASP-2 \citep{albrecht2011}. For reasons of consistency we use the RV based solution of \citet{tripathi2010} for our intial conditions for WASP-3, rather than the more recent tomographical study of \citet{miller2010}. The full set of planets sample is listed in Table\,\ref{tab:misalign}.

\begin{table*}
	\caption{Relevant data for our new misalignment criterion, for a sample of WASP planets with existing Rossiter-McLaughlin measurements. $\lambda$ values are those obtained from our new MCMC analyses. BIC values were calculated from the spectroscopic $\chi^2$ values, using the number of in-transit RV measurements only. Our new misalignment criterion defines systems with a BIC ratio $B\geq1.01$ as misaligned, those with $B\leq0.99$ as aligned, and those with $0.99<B<1.01$ as of indeterminate status.}
	\label{tab:misalign}
	\begin{tabular}{llllllll}
		\hline \\
		System & reference & $\lambda$/$^{\circ}$ & $v\sin I$/km\,s$^{-1}$ & BIC & $\mbox{BIC}_{\mbox{align}}$ & $\Delta\mbox{BIC}$ & $B$ \\ [2pt]
		\hline \\
		WASP-1 & \citet{albrecht2011} & $60.2^{+23.3}_{-126.6}$ & $1.3\pm0.5$ & $255.2\pm22.6$ & $256.7\pm22.7$ & $1.5$ & $1.006$ \\[2pt]
		WASP-3 & \citet{tripathi2010} & $37.9^{+9.3}_{-11.8}$ & $12.9^{+1.1}_{-0.8}$ & $294.7\pm24.3$ & $308.0\pm24.8$ & $13.3$ & $1.045$ \\[2pt]
		WASP-4 & \citet{triaud2010} & $42.0^{+14.3}_{-75.6}$ & $2.5^{+0.4}_{-0.3}$ & $86.8\pm13.2$ & $91.3\pm13.5$ & $4.5$ & $1.052$ \\[2pt]
		WASP-5 & \citet{triaud2010} & $26.2^{+8.1}_{-6.8}$ & $3.5\pm0.2$ & $186.0\pm19.3$ & $199.2\pm20.0$ & $12.1$ & $1.071$ \\[2pt]
		WASP-6 & \citet{gillon2009} & $-7.5^{+20.9}_{-19.1}$ & $1.7^{+0.3}_{-0.2}$ & $134.7\pm16.4$ & $132.0\pm16.2$ & $-2.7$ & $0.980$ \\[2pt]
		WASP-7 & \citet{albrecht2012} & $85.0^{+9.4}_{-8.0}$  & $26.3^{+1.3}_{-1.2}$ & $285.8\pm23.9$ & $451.7\pm30.1$ & $165.9$ & $1.580$ \\[2pt]
		WASP-8 & \citet{queloz2010} & $-106.7^{+3.0}_{-3.5}$ & $2.8^{+0.4}_{-0.3}$ & $380.3\pm27.8$ & $1092.5\pm46.7$ & $712.2$ & $2.873$ \\[2pt]
		WASP-14 & \citet{joshi2009} & $-28.0^{+5.0}_{-5.5}$ & $2.8\pm0.3$ & $171.2\pm18.5$ & $193.5\pm19.7$ & $22.3$ & $1.130$ \\[2pt]
		WASP-15 & \citet{triaud2010} & $-133.8^{+11.7}_{-9.5}$ & $4.5^{+0.4}_{-0.3}$ & $154.4\pm17.6$ & $555.7\pm33.3$ & $401.3$ & $3.599$ \\[2pt]
		WASP-17 & \citet{triaud2010} & $-134.5^{+5.3}_{-7.1}$ & $9.8\pm0.3$ & $342.7\pm26.2$ & $986.9\pm44.4$ & $644.2$ & $2.880$ \\[2pt]
		WASP-18 & \citet{triaud2010} & $20.5^{+10.5}_{-11.5}$ & $12.9\pm0.3$ & $118.8\pm15.4$ & $119.0\pm15.4$ & $0.2$ & $1.002$ \\[2pt]
		WASP-19 & \citet{hellier2011} & $-1.6^{+5.6}_{-5.4}$ & $3.2\pm0.2$ & $81.5\pm12.8$ & $79.8\pm12.6$ & $-1.7$ & $0.979$ \\[2pt]
		WASP-24 & \citet{simpson2011} & $-6.9^{+5.4}_{-5.8}$ & $5.1^{+0.4}_{-0.3}$ & $123.1\pm15.7$ & $119.8\pm15.5$ & $-3.3$ & $0.973$ \\[2pt]
		WASP-38 & \citet{simpson2011} & $-6.1^{+3.3}_{-38.7}$ & $8.2\pm0.3$ & $241.3\pm22.0$ & $240.8\pm21.9$ & $-0.5$ & $0.998$ \\[2pt]
		\hline \\[2pt]
		WASP-16 & this study & $-4.2^{+11.0}_{-13.9}$ & $1.2^{+0.4}_{-0.5}$ & $115.6\pm15.2$ & $112.1\pm15.0$ & $-3.5$ & $0.970$ \\[2pt]
		WASP-25 & this study & $14.6\pm6.7$ & $2.9\pm0.3$ & $116.5\pm15.3$ & $117.5\pm15.3$ & $0.8$ & $1.009$ \\[2pt]
		WASP-31 & this study & $2.8\pm3.1$ & $7.5\pm0.7$ & $73.7\pm12.1$ & $72.2\pm12.0$ & $-1.5$ & $0.980$ \\[2pt]
		\hline \\
	\end{tabular}
\end{table*}

Our test is based on the Bayesian Information Criterion (BIC) \citep{liddle2007},
\begin{equation}
	\mbox{BIC}=\chi^2_{RV} + k\ln(n),
	\label{eq:BIC}
\end{equation}where $k$ is the number of parameters and $n$ is the number of data. Changing the value of $\lambda$ only affects the form of the model RV curve in-transit; we therefore just consider those RV points that lie within a region of the RV curve around phase $0$ defined by the fractional transit width when computing the second term of the BIC. The number of parameters changes according to the choice of priors applied to the MCMC run; adding a long-term RV trend, fitting the RM effect, and allowing the eccentricity to float all add one or more additional parameters to the model.

We carry out two MCMC analyses for each of the systems in our sample, using the same combination of priors for both. The first analysis allows both $\sqrt{v\sin I}\cos\lambda$ and $\sqrt{v\sin I}\sin\lambda$ to float, whilst the second forces an aligned orbit by fixing $\sqrt{v\sin I}\sin\lambda=0$. We calculate the BIC for both runs, before calculating $B=\mbox{BIC}_{\mbox{align}}/\mbox{BIC}$. For the 3 systems presented herein we use our adopted solutions, and carry out an additional run to provide the aligned case. We plot the results for all of the systems as a function of the sky-projected alignment angle.

\begin{figure*}
	\subfloat{
		\includegraphics[width=0.48\textwidth]{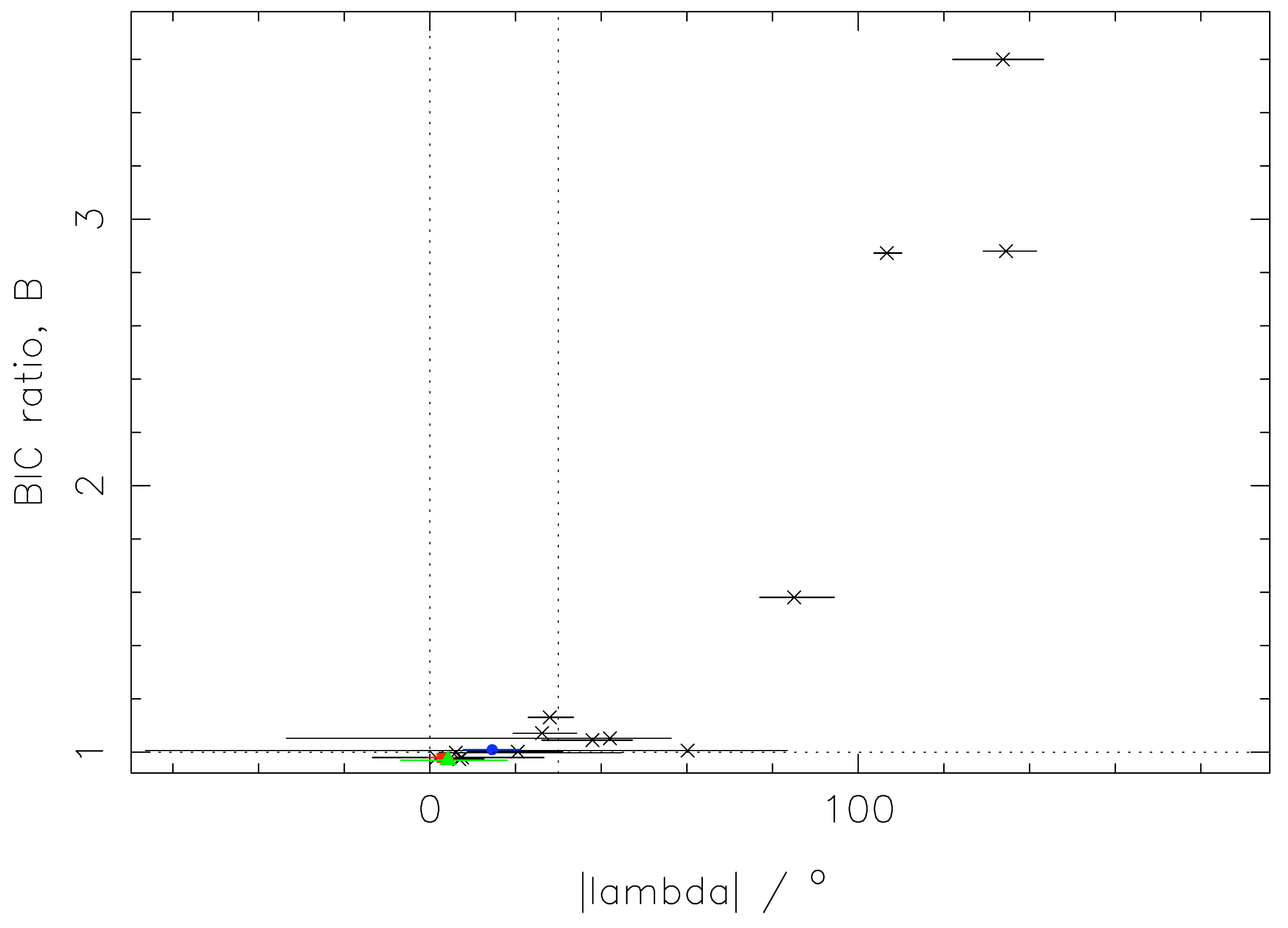}
		\label{fig:BICall}}
	\subfloat{
		\includegraphics[width=0.48\textwidth]{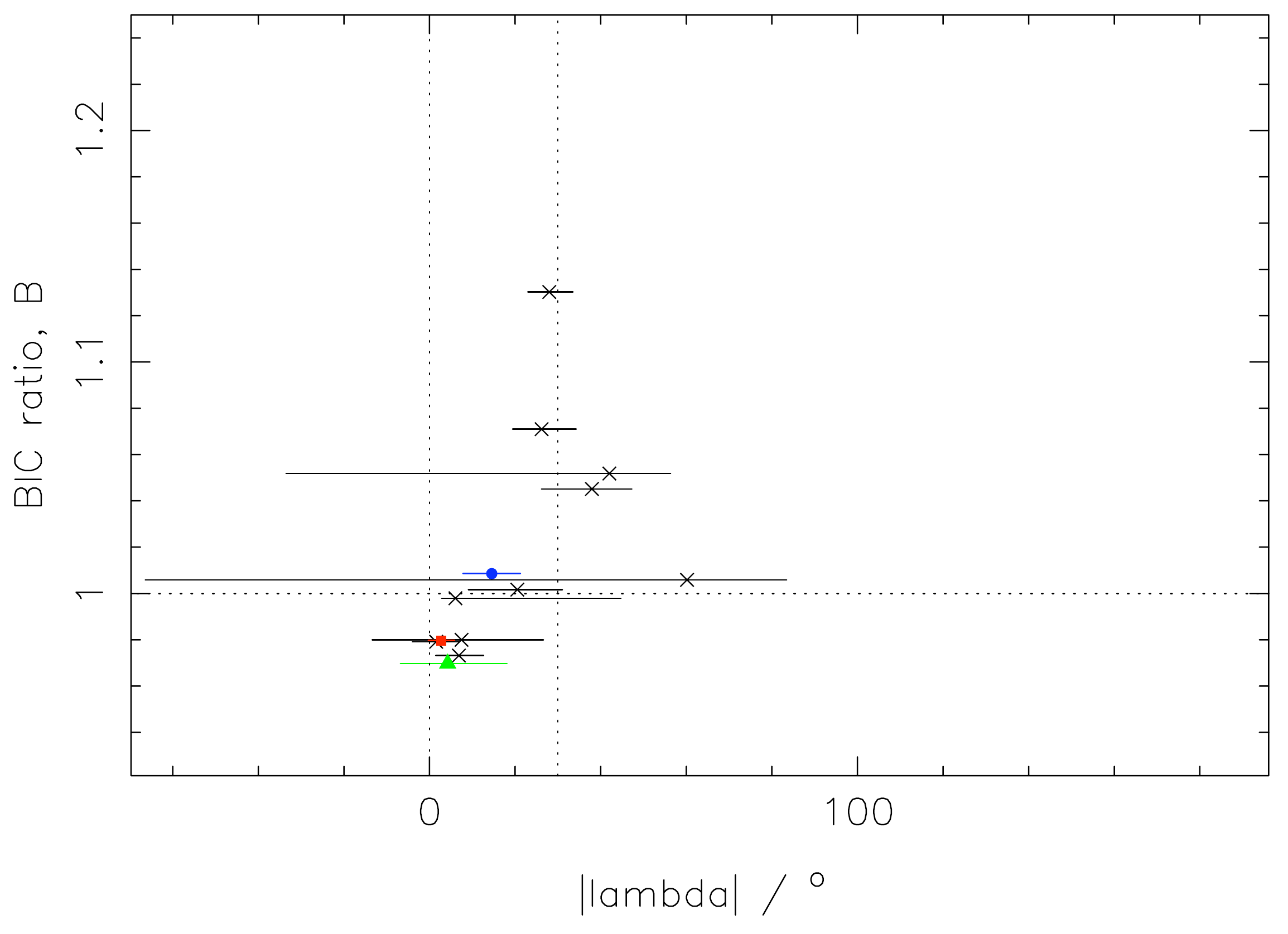}
		\label{fig:BICzoom}}
	\caption{$B$ as a function of $\lambda$ for the sample of planets in Table\,\ref{tab:misalign}, as well as the systems presented in this study. WASP-16 is denoted by a filled, green triangle. WASP-25 is denoted by a filled, blue circle. WASP-31 is denoted by a filled, red square. The horizontal dotted line marks $B=1.00$. The two vertical dotted lines denote $|\lambda|=0^{\circ}$ and $|\lambda|=30^{\circ}$, the existing criterion for misalignment. \textit{Left: } All data. \textit{Right: } A close-up of the heavily populated region in the lower left of the plot, around $B=1.00$ and $|\lambda|=0^{\circ}$. This shows the separation of the systems into several distinct groupings, which lead us to define three categories of alignment. This changes the existing classification of some systems.}
	\label{fig:BIC}
\end{figure*}

We find several distinct groups of systems within our results, which lead us to define three categories of alignment into which systems with RM measurements can be classified. Five systems, including WASP-16 and WASP-31, were found to have$B\leq0.980$, implying that the model with $\lambda=0$ provides a better fit that the free-floating $\lambda$ model. Of these five systems, all would be classified as aligned according to either of the existing misalignment criteria. A further four systems, including WASP-25, are clustered around $B=1.00$, forming a distinct group in figure\,\ref{fig:BICzoom}. Forcing an aligned orbit would seem to make little difference to the quality of the fit between data and model in these cases. Of these systems three would clearly be classed as aligned according to \citetalias{winn2010} and \citetalias{triaud2010}, but the fourth (WASP-1) would actually be classed as misaligned according to \citetalias{winn2010}. The remaining systems clearly lie distinct from those discussed so far, and many are clearly classifiable as misaligned, with $|\lambda|>100^{\circ}$ and $B>1.5$.

In light of these results, we define three categories of alignment. Systems for which $B\leq0.99$ we classify as misaligned. Those with $B\geq1.01$ we classify as aligned. Systems falling between these categories, with $0.99<B<1.01$ we classify as of indeterminate alignment. We would also define a fourth category, that of `no detection', as containing those systems with $v\sin I$ consistent with $0$ to within $1\sigma$, but our current sample contains no systems that meet this requirement.

Some of the systems in Table\,\ref{tab:misalign} warrant a little more examination. WASP-16, despite the relatively poor quality of the RM data that we obtained, can be more strongly considered aligned than WASP-31 with its high quality data. This is an interesting, if puzzling result, but does provide further evidence to support our previous conclusion of an aligned system for WASP-16. WASP-25 is classified as undetermined under our new criteria, possibly owing to the relatively poor match between the shape of the RM anomaly and the best-fitting model. However we note that it lies very close to the boundary between the `undetermined' and `misaligned' categories. Our new MCMC runs for WASP-1 and WASP-4 produce very large error bars on $\lambda$, but end up in different categories despite both failing the \citetalias{triaud2010} criterion of misalignment. Examining their respective RM anomalies we note that both have very low amplitudes, but that the data for WASP-4 is of significantly better quality than that for WASP-1. It is likely that this is responsible for the difference in classification. In addition, \citetalias{triaud2010} noted a substantial correlation between $\lambda$ and $v\sin I$ for WASP-4, arising due to the low impact parameter, which may be producing the large lower error. WASP-38 also exhibits a significant error bar on $\lambda$, and we again note that the RM data is again of somewhat poor quality. New observations of WASP-38 using HARPS may help to improve the quality of the results for the system, allowing us to draw firmer conclusions (Brown et al, \textit{in prep.}).

\section{Conclusions}
\label{sec:conclude}
We have presented analysis of the Rossiter-McLaughlin effect for WASP-16, WASP-25 and WASP-31. We find WASP-16 to have a very low amplitude signal, but the use of two complete spectroscopic transits has enabled us to determine a sky-projected alignment angle of $\lambda=-4.2^{\circ\,+11.0}_{\,\,-13.9}$. For WASP-25 we find a mildly asymmetric RM anomaly with $\lambda=14.6^{\circ}\pm6.7$, and for WASP-31 we obtain $\lambda=2.8^{\circ}\pm3.1$, indicating a well-aligned system.

Since WASP-31 lies so close to the effective temperature that divides the classes of `hot' and `cool' planet hosts, we cannot say how its alignment affects the pattern proposed by \citetalias{winn2010}. WASP-25 on the other hand at first appears to strengthen their hypothesis, with the existing misalignment criteria of both \citetalias{winn2010} and \citetalias{triaud2010} labelling it aligned. We have also presented a new method for determining the alignment or otherwise of an exoplanetary orbit. Our test is based on the BIC statistic, and bases the misalignment or alignment of a system on the ratio of the values of the BIC for the free $\lambda$ case and the aligned case. We classify systems with $B\geq1.01$ as misaligned, those with $B\leq0.99$ as aligned, and those with $0.99\leq B \leq1.01$ as of indeterminate classification. WASP-25 falls in this last category, albeit very close to the boundary with the `misaligned' classification.

The results presented herein bring the analysis of the ensemble of systems with confirmed stellar obliquities up to date. Our results have done little to change the overall picture presented by \citetalias{triaud2010}, instead strengthening the agreement with theoretical predictions for the distributions of both the projected and true stellar obliquities. We should not be too hasty to assume that we have solved the problem of hot Jupiter migration however; new discoveries are constantly causing us to re-evaluate our current understanding.

\section*{Acknowledgments}
\label{sec:acknowledge}
The authors would like to thank the referee, Josh Winn, for his insightful comments and constructive suggestions. D. J. A. Brown would also like to thank Teryuki Hirano for assistance with improving the RM modelling, and Simon Albrecht for helpful discussions regarding analysis of these systems. M. Gillon is FNRS Research Associate. The WASP Consortium consists of representatives from the Universities of Keele, Leicester, The Open University, Queens University Belfast and St Andrews, along with the Isaac Newton Group (La Palma) and the Instituto de Astrofisca de Canarias (Tenerife). The SuperWASP and WASP-S cameras were constructed and operated with funds made available from Consortium Universities and PPARC/STFC. This research has made use of NASA's Astrophysics Data System Bibliographic Services, and the ArXiv preprint service hosted by Cornell University.

\bibliographystyle{mn2e}
%\bibliography{swasp_transrch}

\bsp

%\newpage
\appendix
\newpage

\section{Additional results}
\label{sec:moreres}

\begin{table*}
	\caption{Parameters from the best-fitting, adopted models for the three WASP planetary systems studied in the main text}
	\label{tab:results}
	\begin{tabular}{llllllll}
		\hline \\
		System & $v\sin I$ prior & MS prior & $\dot{\gamma}$/ms$^{-1}$yr$^{-1}$ & eccentricity & jitter /m\,s$^{-1}$ & $v\sin I$/km\,s$^{-1}$ & $\lambda$/$^{\circ}$ \\
		\hline \\
		WASP-16 & off & off & $0$ & $0$ & $1.0$ & $1.2^{+0.4}_{-0.5}$ & $-4.2^{+11.0}_{-13.9}$ \\[2pt]
		WASP-25 & off & on & $0$ & $0$ & $3.6$ & $2.9\pm0.3$ & $14.6\pm6.7$ \\[2pt]
		WASP-31 & off & on & $0$ & $0$ & $1.0$ & $7.5\pm0.7$ & $2.8\pm3.1$ \\[2pt]
		\hline \\
	\end{tabular}
\end{table*}

\section{Journal of observations}
\label{sec:journal}
\begin{table}
	\caption{Radial velocity data for WASP-16 obtained using the CORALIE high precision {\'e}chelle spectrograph.}
	\label{tab:W16coralie}
	\begin{tabular}{llll}
		\hline \\
		HJD(-2450000) & $T{\rm exp}$/s & RV/km\,s$^{-1}$ & $\sigma_{RV}$/km\,s$^{-1}$ \\
		\hline \\
		$4535.864842$ & $900$ & $-1.99772$ & $0.01591$ \\
		$4537.849158$ & $1427$ & $-1.96688$ & $0.00853$ \\
		$4538.858364$ & $1800$ & $-2.00734$ & $0.00899$ \\
		$4558.780835$ & $1800$ & $-1.83336$ & $0.00723$ \\
		$4560.709473$ & $1800$ & $-2.00513$ & $0.00725$ \\
		$4561.688137$ & $1800$ & $-1.82730$ & $0.00785$ \\
		$4589.705102$ & $1800$ & $-1.84255$ & $0.00875$ \\
		$4591.706755$ & $1800$ & $-2.03571$ & $0.00892$ \\
		$4652.495906$ & $1800$ & $-1.82493$ & $0.00808$ \\
		$4656.551645$ & $1800$ & $-2.02421$ & $0.00787$ \\
		$4657.577293$ & $1800$ & $-1.96640$ & $0.00957$ \\
		$4663.539741$ & $1800$ & $-2.02961$ & $0.00969$ \\
		$4664.616769$ & $1800$ & $-1.78590$ & $0.01108$ \\
		$4682.521501$ & $1800$ & $-1.98118$ & $0.00754$ \\
		$4881.869213$ & $1800$ & $-2.02245$ & $0.00813$ \\
		$4882.801025$ & $1800$ & $-1.83289$ & $0.00823$ \\
		$4884.737094$ & $1800$ & $-2.04565$ & $0.00778$ \\
		$4891.805707$ & $1800$ & $-1.90043$ & $0.00798$ \\
		$4892.723980$ & $1800$ & $-1.83413$ & $0.00891$ \\
		$4941.728231$ & $1800$ & $-1.88737$ & $0.00748$ \\
		$4943.730102$ & $1800$ & $-2.04677$ & $0.00753$ \\
		$4944.739293$ & $1800$ & $-1.91359$ & $0.00860$ \\
		$4945.799895$ & $1800$ & $-1.85815$ & $0.00807$ \\
		$4947.745317$ & $1800$ & $-1.93960$ & $0.00741$ \\
		$4948.673112$ & $1800$ & $-1.82992$ & $0.00743$ \\
		$4972.707323$ & $1800$ & $-1.93123$ & $0.00854$ \\
		$4975.733486$ & $1800$ & $-1.93144$ & $0.01100$ \\
		$4982.647535$ & $1800$ & $-1.83433$ & $0.01036$ \\
		$4984.642389$ & $1800$ & $-2.04210$ & $0.00892$ \\
		$4985.694776$ & $1800$ & $-1.81561$ & $0.00802$ \\
		$5391.544362$ & $1800$ & $-1.80313$ & $0.00867$ \\
		\hline \\
	\end{tabular}
\end{table}

\begin{table}
	\caption{Radial velocity data for WASP-16, for the first transit obtained using the HARPS high precision {\'e}chelle spectrograph on the night of 2010 March 21.}
	\label{tab:W16harps}
	\begin{tabular}{llll}
		\hline \\
		HJD(-2450000) & $T{\rm exp}$/s & RV/km\,s$^{-1}$ & $\sigma_{RV}$/km\,s$^{-1}$ \\
		\hline \\
		$5275.661171$ & $1800$ & $-1.80610$ & $0.00337$ \\
		$5275.907691$ & $1800$ & $-1.78144$ & $0.00188$ \\
		$5276.661941$ & $500$ & $-1.88264$ & $0.00533$ \\
		$5276.668446$ & $500$ & $-1.89263$ & $0.00546$ \\
		$5276.674824$ & $500$ & $-1.88914$ & $0.00583$ \\
		$5276.681375$ & $500$ & $-1.87845$ & $0.00547$ \\
		$5276.687753$ & $500$ & $-1.89900$ & $0.00557$ \\
		$5276.694258$ & $500$ & $-1.88947$ & $0.00569$ \\
		$5276.700693$ & $500$ & $-1.89868$ & $0.00555$ \\
		$5276.707094$ & $500$ & $-1.89750$ & $0.00552$ \\
		$5276.713599$ & $500$ & $-1.88945$ & $0.00586$ \\
		$5276.720046$ & $500$ & $-1.91121$ & $0.00614$ \\
		$5276.726493$ & $500$ & $-1.89650$ & $0.00623$ \\
		$5276.732929$ & $500$ & $-1.88385$ & $0.00657$ \\
		$5276.739376$ & $500$ & $-1.90596$ & $0.00640$ \\
		$5276.745812$ & $500$ & $-1.90686$ & $0.00704$ \\
		$5276.752143$ & $500$ & $-1.90101$ & $0.00698$ \\
		$5276.758579$ & $500$ & $-1.91736$ & $0.00742$ \\
		$5276.765605$ & $500$ & $-1.90950$ & $0.00627$ \\
		$5276.771589$ & $500$ & $-1.91143$ & $0.00447$ \\
		$5276.778140$ & $500$ & $-1.91692$ & $0.00439$ \\
		$5276.784344$ & $500$ & $-1.91573$ & $0.00479$ \\
		$5276.790838$ & $500$ & $-1.92779$ & $0.00526$ \\
		$5276.797459$ & $500$ & $-1.92243$ & $0.00510$ \\
		$5276.803964$ & $500$ & $-1.90902$ & $0.00433$ \\
		$5276.810411$ & $500$ & $-1.92567$ & $0.00391$ \\
		$5276.816441$ & $500$ & $-1.91540$ & $0.00416$ \\
		$5276.823178$ & $500$ & $-1.92366$ & $0.00424$ \\
		$5276.829336$ & $500$ & $-1.92742$ & $0.00454$ \\
		$5276.835887$ & $500$ & $-1.92693$ & $0.00522$ \\
		$5276.842334$ & $500$ & $-1.92962$ & $0.00572$ \\
		$5276.848723$ & $500$ & $-1.94183$ & $0.00699$ \\
		$5276.855228$ & $500$ & $-1.94543$ & $0.00926$ \\
		$5276.861907$ & $500$ & $-1.92508$ & $0.00830$ \\
		$5277.630948$ & $1800$ & $-2.02847$ & $0.00222$ \\
		$5277.861599$ & $1800$ & $-1.99854$ & $0.00196$ \\
		$5278.632376$ & $1800$ & $-1.82733$ & $0.00398$ \\
		$5278.857922$ & $1800$ & $-1.79546$ & $0.00208$ \\
		$5279.627285$ & $1800$ & $-1.84379$ & $0.00264$ \\
		$5279.913540$ & $1500$ & $-1.91554$ & $0.00242$ \\
		$5280.624797$ & $1800$ & $-2.03079$ & $0.00266$ \\
		$5280.916481$ & $1200$ & $-2.00824$ & $0.00283$ \\
		\hline \\
	\end{tabular}
\end{table}

\begin{table}
	\caption{Radial velocity data for WASP-16, for the second transit obtained using the HARPS high precision {\'e}chelle spectrograph on the night of 2011 May 12.}
	\label{tab:W16harps2}
	\begin{tabular}{llll}
		\hline \\
		HJD(-2450000) & $T{\rm exp}$/s & RV/km\,s$^{-1}$ & $\sigma_{RV}$/km\,s$^{-1}$ \\
		\hline \\
		$5685.845943$ & $900$ & $-2.02724$ & $0.00305$ \\
		$5687.838150$ & $900$ & $-1.79259$ & $0.00383$ \\
		$5692.662149$ & $900$ & $-1.99841$ & $0.00380$ \\
		$5692.796210$ & $900$ & $-1.96847$ & $0.00309$ \\
		$5693.517817$ & $900$ & $-1.81013$ & $0.00298$ \\
		$5693.800775$ & $900$ & $-1.78196$ & $0.00285$ \\
		$5694.581176$ & $600$ & $-1.88349$ & $0.00344$ \\
		$5694.588340$ & $600$ & $-1.88597$ & $0.00302$ \\
		$5694.595389$ & $600$ & $-1.88670$ & $0.00310$ \\
		$5694.602900$ & $600$ & $-1.88871$ & $0.00305$ \\
		$5694.610180$ & $600$ & $-1.89619$ & $0.00290$ \\
		$5694.616904$ & $500$ & $-1.89308$ & $0.00323$ \\
		$5694.623386$ & $500$ & $-1.88547$ & $0.00323$ \\
		$5694.629531$ & $500$ & $-1.89159$ & $0.00309$ \\
		$5694.635631$ & $500$ & $-1.89935$ & $0.00312$ \\
		$5694.641904$ & $500$ & $-1.89371$ & $0.00307$ \\
		$5694.648003$ & $500$ & $-1.89694$ & $0.00318$ \\
		$5694.654149$ & $500$ & $-1.91191$ & $0.00298$ \\
		$5694.660364$ & $500$ & $-1.91005$ & $0.00311$ \\
		$5694.666406$ & $500$ & $-1.91009$ & $0.00300$ \\
		$5694.672609$ & $500$ & $-1.91343$ & $0.00311$ \\
		$5694.678824$ & $500$ & $-1.91547$ & $0.00327$ \\
		$5694.684924$ & $500$ & $-1.91384$ & $0.00313$ \\
		$5694.691070$ & $500$ & $-1.91484$ & $0.00330$ \\
		$5694.697227$ & $500$ & $-1.91701$ & $0.00308$ \\
		$5694.703373$ & $500$ & $-1.91926$ & $0.00308$ \\
		$5694.709460$ & $500$ & $-1.91211$ & $0.00335$ \\
		$5694.715664$ & $500$ & $-1.91534$ & $0.00353$ \\
		$5694.721821$ & $500$ & $-1.91948$ & $0.00349$ \\
		$5694.727979$ & $500$ & $-1.92566$ & $0.00346$ \\
		$5694.734078$ & $500$ & $-1.92109$ & $0.00367$ \\
		$5694.740351$ & $500$ & $-1.92435$ & $0.00377$ \\
		$5694.746612$ & $500$ & $-1.91926$ & $0.00364$ \\
		$5694.752596$ & $500$ & $-1.93024$ & $0.00354$ \\
		$5695.501446$ & $900$ & $-2.03120$ & $0.00281$ \\
		\hline \\
	\end{tabular}
\end{table}

\begin{table}
	\caption{Radial velocity data for WASP-25 obtained using the CORALIE high precision {\'e}chelle spectrograph.}
	\label{tab:W25coralie}
	\begin{tabular}{llll}
		\hline \\
		HJD(-2450000) & $t{\rm exp}$/s & RV/km\,s$^{-1}$ & $\sigma_{RV}$/km\,s$^{-1}$ \\
		\hline \\
		$4829.822664$ & $1800$ & $-2.57717$ & $0.01282$ \\
		$4896.769798$ & $1800$ & $-2.65105$ & $0.01069$ \\
		$4940.709168$ & $1800$ & $-2.71589$ & $0.01154$ \\
		$4941.704336$ & $1800$ & $-2.61855$ & $0.01153$ \\
		$4942.725717$ & $1800$ & $-2.57632$ & $0.01238$ \\
		$4943.637434$ & $1800$ & $-2.61828$ & $0.01246$ \\
		$4944.715466$ & $1800$ & $-2.67966$ & $0.01207$ \\
		$4945.726530$ & $1800$ & $-2.61467$ & $0.01305$ \\
		$4946.616622$ & $1800$ & $-2.58169$ & $0.01266$ \\
		$4947.601618$ & $1800$ & $-2.64133$ & $0.01096$ \\
		$4947.791245$ & $1800$ & $-2.68927$ & $0.01347$ \\
		$4948.613002$ & $1800$ & $-2.70418$ & $0.01098$ \\
		$4949.803142$ & $560$ & $-2.55132$ & $0.01819$ \\
		$4950.622083$ & $1800$ & $-2.59141$ & $0.01348$ \\
		$4951.695324$ & $1800$ & $-2.70149$ & $0.01218$ \\
		$4971.645302$ & $1800$ & $-2.67821$ & $0.02101$ \\
		$4972.672436$ & $1800$ & $-2.56129$ & $0.01319$ \\
		$4973.515713$ & $1800$ & $-2.58676$ & $0.01269$ \\
		$4974.678659$ & $1800$ & $-2.71413$ & $0.01359$ \\
		$4975.537940$ & $1800$ & $-2.66695$ & $0.01384$ \\
		$4976.683662$ & $1800$ & $-2.55567$ & $0.01304$ \\
		$4982.619435$ & $1800$ & $-2.66448$ & $0.02096$ \\
		$4983.621314$ & $1800$ & $-2.56777$ & $0.01486$ \\
		$4983.644577$ & $1800$ & $-2.59698$ & $0.01450$ \\
		$4984.578450$ & $1800$ & $-2.55837$ & $0.01474$ \\
		$4985.609967$ & $1800$ & $-2.69905$ & $0.01189$ \\
		$4995.555496$ & $1800$ & $-2.50858$ & $0.01396$ \\
		$5009.628712$ & $1800$ & $-2.60564$ & $0.01823$ \\
		$5010.596729$ & $1800$ & $-2.53871$ & $0.02313$ \\
		\hline \\
	\end{tabular}
\end{table}

\begin{table}
	\caption{Radial velocity data for WASP-25 obtained using the HARPS high precision {\'e}chelle spectrograph. The point denoted by $^*$ was omitted from the analysis (see text for details).}
	\label{tab:W25harps}
	\begin{tabular}{llll}
		\hline \\
		HJD(-2450000) & $t{\rm exp}$/s & RV/km\,s$^{-1}$ & $\sigma_{RV}$/km\,s$^{-1}$ \\
		\hline \\
		$5296.540546$ & $1200$ & $-2.54661$ & $0.00329$ \\
		$5296.635060$ & $1200$ & $-2.54381$ & $0.00398$ \\
		$5297.506446$ & $1200$ & $-2.61464$ & $0.00509$ \\
		$5297.518749$ & $400$ & $-2.62531$ & $0.01007$ \\
		$5297.523714$ & $400$ & $-2.63325$ & $0.00942$ \\
		$5297.528714$ & $400$ & $-2.60898$ & $0.01012$ \\
		$5297.533807$ & $400$ & $-2.61031$ & $0.01041$ \\
		$5297.538714$ & $400$ & $-2.60809$ & $0.00999$ \\
		$5297.543714$ & $400$ & $-2.59973$ & $0.01056$ \\
		$5297.548668$ & $400$ & $-2.59327$ & $0.01108$ \\
		$5297.553761$ & $400$ & $-2.58984$ & $0.01028$ \\
		$5297.559131$ & $400$ & $-2.61140$ & $0.01786$ \\
		$5297.563761$ & $400$ & $-2.60108$ & $0.01177$ \\
		$5297.568668$ & $400$ & $-2.60927$ & $0.01087$ \\
		$5297.573715$ & $400$ & $-2.60539$ & $0.01075$ \\
		$5297.578761$ & $400$ & $-2.62964$ & $0.01153$ \\
		$5297.583668$ & $400$ & $-2.62703$ & $0.01118$ \\
		$5297.588669$ & $400$ & $-2.61584$ & $0.01225$ \\
		$5297.593715$ & $400$ & $-2.64141$ & $0.01190$ \\
		$5297.598669$ & $400$ & $-2.64658$ & $0.01232$ \\
		$5297.603715$ & $400$ & $-2.65755$ & $0.01234$ \\
		$5297.608715$ & $400$ & $-2.67520$ & $0.01246$ \\
		$5297.613761$ & $400$ & $-2.67558$ & $0.01254$ \\
		$5297.618715$ & $400$ & $-2.68567$ & $0.01244$ \\
		$5297.623669^*$ & $400$ & $-2.63635$ & $0.01215$ \\
		$5297.628854$ & $400$ & $-2.67450$ & $0.01065$ \\
		$5297.633761$ & $400$ & $-2.65389$ & $0.00837$ \\
		$5297.638715$ & $400$ & $-2.63022$ & $0.00840$ \\
		$5297.643773$ & $400$ & $-2.63126$ & $0.00885$ \\
		$5297.648727$ & $400$ & $-2.61768$ & $0.00871$ \\
		$5297.653727$ & $400$ & $-2.63157$ & $0.00862$ \\
		$5297.658681$ & $400$ & $-2.63982$ & $0.00841$ \\
		$5297.663773$ & $400$ & $-2.62371$ & $0.00834$ \\
		$5297.668773$ & $400$ & $-2.63776$ & $0.00800$ \\
		$5297.673727$ & $400$ & $-2.64716$ & $0.00811$ \\
		$5297.678727$ & $400$ & $-2.63753$ & $0.00780$ \\
		$5297.683681$ & $400$ & $-2.63915$ & $0.00781$ \\
		$5297.688820$ & $400$ & $-2.63818$ & $0.00823$ \\
		$5297.693727$ & $400$ & $-2.64295$ & $0.00763$ \\
		$5297.698774$ & $400$ & $-2.62870$ & $0.00772$ \\
		$5297.703635$ & $400$ & $-2.63667$ & $0.00757$ \\
		$5297.708727$ & $400$ & $-2.63667$ & $0.00781$ \\
		$5297.713727$ & $400$ & $-2.63031$ & $0.00803$ \\
		$5297.718681$ & $400$ & $-2.65168$ & $0.00756$ \\
		$5297.723774$ & $400$ & $-2.64533$ & $0.00785$ \\
		$5297.833578$ & $1200$ & $-2.65676$ & $0.00352$ \\
		$5298.535157$ & $1200$ & $-2.69608$ & $0.00406$ \\
		$5298.716015$ & $1200$ & $-2.69119$ & $0.00285$ \\
		$5298.830796$ & $1200$ & $-2.69107$ & $0.00287$ \\
		$5299.544943$ & $1200$ & $-2.60603$ & $0.00327$ \\
		$5299.701761$ & $1200$ & $-2.62922$ & $0.01842$ \\
		$5299.838220$ & $1384$ & $-2.57573$ & $0.01224$ \\
		\hline \\
	\end{tabular}
\end{table}

\begin{table}
	\caption{Radial velocity data for WASP-31 obtained using the CORALIE high precision {\'e}chelle spectrograph.}
	\label{tab:W31coralie}
	\begin{tabular}{llll}
		\hline \\
		HJD(-2450000) & $t{\rm exp}$/s & RV/km\,s$^{-1}$ & $\sigma_{RV}$/km\,s$^{-1}$ \\
		\hline \\
		$4835.809755$ & $1800$ & $-0.20260$ & $0.02945$ \\
		$4837.773728$ & $1800$ & $-0.08457$ & $0.03163$ \\
		$4840.765776$ & $1800$ & $-0.07974$ & $0.03752$ \\
		$4880.767231$ & $1800$ & $-0.20651$ & $0.03722$ \\
		$4939.627676$ & $1800$ & $-0.07528$ & $0.03471$ \\
		$4941.567460$ & $1800$ & $-0.19372$ & $0.02721$ \\
		$4942.654757$ & $1800$ & $-0.12324$ & $0.02598$ \\
		$4943.610624$ & $1800$ & $-0.07939$ & $0.03435$ \\
		$4944.555415$ & $1800$ & $-0.16578$ & $0.02620$ \\
		$4945.544750$ & $1800$ & $-0.16158$ & $0.03220$ \\
		$4946.591778$ & $1800$ & $-0.04422$ & $0.03441$ \\
		$4947.555094$ & $1800$ & $-0.14075$ & $0.02656$ \\
		$4948.588069$ & $1800$ & $-0.18274$ & $0.02861$ \\
		$4950.597039$ & $1800$ & $-0.08667$ & $0.03009$ \\
		$4951.608264$ & $1800$ & $-0.22821$ & $0.03366$ \\
		$4971.548671$ & $1800$ & $-0.02806$ & $0.07763$ \\
		$4973.489951$ & $1800$ & $-0.09189$ & $0.03084$ \\
		$4974.608541$ & $1800$ & $-0.02130$ & $0.03225$ \\
		$4975.511060$ & $1800$ & $-0.12628$ & $0.03381$ \\
		$4983.595539$ & $1800$ & $-0.10028$ & $0.04415$ \\
		$4984.467744$ & $1800$ & $-0.06112$ & $0.02895$ \\
		$4985.530406$ & $1800$ & $-0.18886$ & $0.03032$ \\
		$4994.508045$ & $1800$ & $-0.09735$ & $0.03237$ \\
		$4994.531307$ & $1800$ & $-0.09484$ & $0.03469$ \\
		$4995.463376$ & $1800$ & $-0.12246$ & $0.03547$ \\
		$4995.486741$ & $1800$ & $-0.20999$ & $0.03214$ \\
		$4996.459605$ & $1800$ & $-0.19034$ & $0.03472$ \\
		$4996.482971$ & $1800$ & $-0.10984$ & $0.03117$ \\
		$4999.536757$ & $1800$ & $-0.21980$ & $0.05522$ \\
		$4999.560099$ & $1800$ & $-0.15670$ & $0.06560$ \\
		$5006.521354$ & $1800$ & $-0.18493$ & $0.04217$ \\
		$5012.492297$ & $1800$ & $-0.09667$ & $0.03739$ \\
		$5013.497045$ & $1800$ & $-0.19082$ & $0.04459$ \\
		$5029.465780$ & $1800$ & $-0.11660$ & $0.05041$ \\
		$5168.846768$ & $1800$ & $-0.15389$ & $0.01852$ \\
		$5203.782854$ & $2700$ & $-0.20152$ & $0.02084$ \\
		$5290.715577$ & $2700$ & $-0.09351$ & $0.01809$ \\
		$5291.699859$ & $2700$ & $-0.12273$ & $0.01821$ \\
		$5293.696819$ & $2700$ & $-0.07455$ & $0.01987$ \\
		$5294.733907$ & $2700$ & $-0.09785$ & $0.01925$ \\
		$5296.704921$ & $2700$ & $-0.13168$ & $0.01772$ \\
		$5298.693406$ & $2700$ & $-0.18210$ & $0.01718$ \\
		$5300.589643$ & $2700$ & $-0.06706$ & $0.02250$ \\
		$5326.628560$ & $2700$ & $-0.13831$ & $0.01945$ \\
		$5327.604475$ & $2700$ & $-0.08093$ & $0.02737$ \\
		$5328.608442$ & $2700$ & $-0.09912$ & $0.01913$ \\
		$5334.544675$ & $2700$ & $-0.05571$ & $0.02129$ \\
		\hline \\
	\end{tabular}
\end{table}

\begin{table}
	\caption{Radial velocity data for WASP-31 obtained using the HARPS high precision {\'e}chelle spectrograph.}
	\label{tab:W31harps}
	\begin{tabular}{llll}
		\hline \\
		HJD(-2450000) & $t{\rm exp}$/s & RV/km\,s$^{-1}$ & $\sigma_{RV}$/km\,s$^{-1}$ \\
		\hline \\
		$5298.496133$ & $1200$ & $-0.15638$ & $0.00896$ \\
		$5298.749145$ & $1200$ & $-0.17218$ & $0.00931$ \\
		$5299.504441$ & $1200$ & $-0.17141$ & $0.01114$ \\
		$5299.716991$ & $1200$ & $-0.29541$ & $0.04800$ \\
		$5300.509357$ & $1200$ & $-0.07910$ & $0.01138$ \\
		$5300.742948$ & $1200$ & $-0.08587$ & $0.01161$ \\
		$5301.582822$ & $1200$ & $-0.11516$ & $0.00934$ \\
		$5301.597544$ & $1200$ & $-0.11250$ & $0.01003$ \\
		$5301.612960$ & $1200$ & $-0.12527$ & $0.00984$ \\
		$5301.627126$ & $900$ & $-0.10841$ & $0.01418$ \\
		$5301.638028$ & $900$ & $-0.09394$ & $0.01141$ \\
		$5301.648803$ & $900$ & $-0.06309$ & $0.01227$ \\
		$5301.659393$ & $900$ & $-0.08776$ & $0.01100$ \\
		$5301.670307$ & $900$ & $-0.11321$ & $0.01136$ \\
		$5301.681082$ & $900$ & $-0.13542$ & $0.01119$ \\
		$5301.691973$ & $900$ & $-0.16717$ & $0.01218$ \\
		$5301.702354$ & $900$ & $-0.19579$ & $0.01297$ \\
		$5301.713349$ & $900$ & $-0.16796$ & $0.01407$ \\
		$5301.723522$ & $900$ & $-0.16142$ & $0.01884$ \\
		$5301.734934$ & $900$ & $-0.14695$ & $0.01771$ \\
		$5301.750477$ & $1200$ & $-0.13378$ & $0.01244$ \\
		$5301.764794$ & $1200$ & $-0.14088$ & $0.01586$ \\
		$5301.780360$ & $1200$ & $-0.15031$ & $0.01738$ \\
		$5305.613529$ & $1200$ & $-0.18166$ & $0.01096$ \\
		$5307.584640$ & $1200$ & $-0.06475$ & $0.00809$ \\
		\hline \\
	\end{tabular}
\end{table}

\label{lastpage}

\end{document}